\documentclass[10pt]{article}

\usepackage{grffile}
\usepackage{ctable}

\usepackage[small]{caption}
\usepackage{graphicx}

\usepackage{amsmath,amsfonts,amssymb}
\usepackage{amsthm}
\usepackage{color}

\usepackage[letterpaper,textheight=9in,textwidth=7in]{geometry}

\usepackage[skip=10pt,font=footnotesize]{caption}
\graphicspath{{}}

 {\begin{trivlist} \item[]{\bf Proof. }}%
 {\hspace*{\fill}$\rule{.4\baselineskip}{.4\baselineskip}$\end{trivlist}}

 {\begin{trivlist}\item[]\textbf{Acknowledgments.}}{\end{trivlist}}

 {\begin{center}\textbf{Abstract}}{\end{center}}

\makeatletter\@addtoreset{figure}{section}\makeatother

\makeatletter \@addtoreset{equation}{section} \makeatother

\def\XXint#1#2#3{{\setbox0=\hbox{$#1{#2#3}{\int}$ }
\vcenter{\hbox{$#2#3$ }}\kern-.6\wd0}}

\newcommand{\R}{\mathbb{R}}
\newcommand{\C}{\mathbb{C}}

\newcommand{\ba}{\begin{align}}
\newcommand{\ea}{\end{align}}

\newcommand{\rmi}{\mathrm{i}}

\newcommand{\rmd}{\mathrm{d}}

\newcommand{\rme}{\mathrm{e}}

\newcommand{\rmO}{\mathcal{O}}
\renewcommand{\Re}{\mathrm{Re}\,}
\renewcommand{\Im}{\mathrm{Im}\,}

\newcommand{\eps}{{\varepsilon}}

\begin{document}
\begin{center}
{\fontsize{15}{15}\fontfamily{cmr}\fontseries{b}\selectfont{Instability in large bounded domains --- branched versus unbranched resonances}}\\[0.2in]
Montie Avery$\,^1$, Cedric Dedina$\,^1$, Aislinn Smith$\,^2$, and Arnd Scheel$\,^1$. \\[0.1in]
\textit{\footnotesize 
$\,^1$University of Minnesota, School of Mathematics,   206 Church St. S.E., Minneapolis, MN 55455, USA\\
$\,^2$University of Texas at Austin, Department of Mathematics, 1 University Station, Austin, TX 78712,  USA}
\date{\small \today} 
\end{center}

\begin{abstract}
We study transitions from convective to absolute instability near a trivial state in large bounded domains for prototypical model problems in the presence of transport and negative nonlinear feedback. We identify two generic scenarios, depending on the nature of the linear mechanism for instability, which both lead to different, universal bifurcation diagrams. In the first, classical case of a linear branched resonance the transition is hard, that is, small changes in a control parameter lead to a  finite-size state. In the second, novel case of an unbranched resonance, the transition is gradual. In both cases, the bifurcation diagram is determined by interaction of the leading edge of an invasion front with upstream boundary conditions. Technically, we analyze this interaction in a heteroclinic gluing bifurcation analysis that uses geometric desingularization of the trivial state. 
%
\end{abstract}

\setlength{\parskip}{4pt}
\setlength{\parindent}{0pt}

\section{Introduction}
Instabilities of simple equilibrium states in spatially extended systems are of interest in many experimental contexts ranging from plasma and fluid dynamics to pattern-forming systems in biology or material science. ``Simple'' here refers to  spatially constant states, such as laminar flows in fluids or unpatterned states in pattern-forming systems. On the one hand, one wishes to determine parameter regimes where such a simple state is experimentally observable. On the other hand, one would like to understand dynamics past the instability when the spatio-temporal growth of disturbances can lead to the selection of new non-trivial nonlinear states.  An important aspect of instabilities is their spatio-temporal behavior, when initiated by localized fluctuations around the unstable state. The growth of such localized perturbations in space and time can be described by fixing a frame of reference and observing the temporal evolution in a finite window of observation. Instabilities are referred to as absolute when growth is observed in this finite window and as convective when perturbations grow in a  translation-invariant norm but not in the fixed finite window of observation. 

\textbf{Unbounded domains.}
Transitions from convective to absolute instabilities can be analyzed in idealized unbounded, linear systems through an analysis of the pointwise Green's function of the linear evolution, obtained by Fourier-Laplace transform. Deforming the Fourier-Laplace  integration contours, one finds pointwise exponential growth rates as determined by pinched double roots of the dispersion relation; see for instance \cite{bers84} or \S \ref{s:ap}. Absolute instability corresponds to pointwise growth rates with positive real part. In nonlinear systems, the spatial spreading of instabilities can often be described by the propagation of nonlinear fronts: the sign of the speed of the nonlinear front in the given frame of reference then discriminates between convective and absolute instabilities; see \cite{vansaarloos03} for a review on front propagation into unstable states. To determine the speed of nonlinear fronts, one usually looks for fronts propagating at a linear spreading speed which is found as the speed of the frame of reference in which the instability is marginal, at the transition from convective to absolute. Stable fronts propagating at this linear speed are referred to as pulled fronts. If fronts propagating at this linear speed are unstable against localized modes, invasion can be faster, mediated by a pushed front. Pushed fronts are usually excluded when nonlinearities give negative feedback, which is the situation that we shall focus on in this work. 

Recently \cite{holzerscheel14,faye17,holzerremnant}, the pinched double root criterion for absolute instability and linear selected front speeds was embedded into a larger, previously unnoticed family of linear criteria for spreading speeds. In fact,  pinched double roots can be viewed as quite particular complex resonances, where two complex spatio-temporal modes $\rme^{\nu_{1,2}x +\lambda t}$ collide, $\nu_1=\nu_2$ at $\lambda=\lambda_\mathrm{br}$, and unfold with $\nu_{1,2}-\nu_\mathrm{br}\sim \pm\sqrt{\lambda-\lambda_\mathrm{br}}$. 
We refer to this classical scenario as a \emph{branched resonance} since spatial modes $\nu$ are analytic on a branched Riemann surface and in 1:1-resonance at $\lambda=\lambda_\mathrm{br}$. It was noted in \cite{holzerscheel14} that $\nu_{1,2}$ may well be genuinely analytic in $\lambda$ (rather than in $\sqrt{\lambda-\lambda_\mathrm{br}}$) at a pinched double root, leading to what we shall refer to as an unbranched 1:1 resonance. More general resonances, for instance a 3-mode resonance where spatio-temporal modes satisfy $\lambda_1=\lambda_2+\lambda_3$ and $\nu_1=\nu_2+\nu_3$, were found in \cite{faye17} to determine linear spreading speeds in situations where, roughly speaking, stationary and oscillatory modes are simultaneously present in instabilities. Unbranched resonances induce associated growth modes only when resonant modes are coupled in the equation through an appropriate linear (for 1:1 resonances) or nonlinear term. We refer to all these resonances, where spatial roots are analytic as functions of $\lambda$, as unbranched resonances. 

We emphasize that in systems without reflection symmetry, the onsets of convective and of absolute instability are not close in parameter space. Nonlinear ``saturated'' stable states that bifurcate at the onset of convective instability and that  are usually selected through front invasion past the onset of absolute instability are therefore typically finite-size, not small near onset, such that ``universal'' amplitude equations do not universally describe such transitions at onset of absolute instability. 

Our objective here is to nevertheless analyze, with some generality and based on conceptual assumptions, the effect of those different linear mechanisms on bifurcation diagrams in large bounded domains.

\begin{figure}[h]
 \includegraphics[height=2.1in]{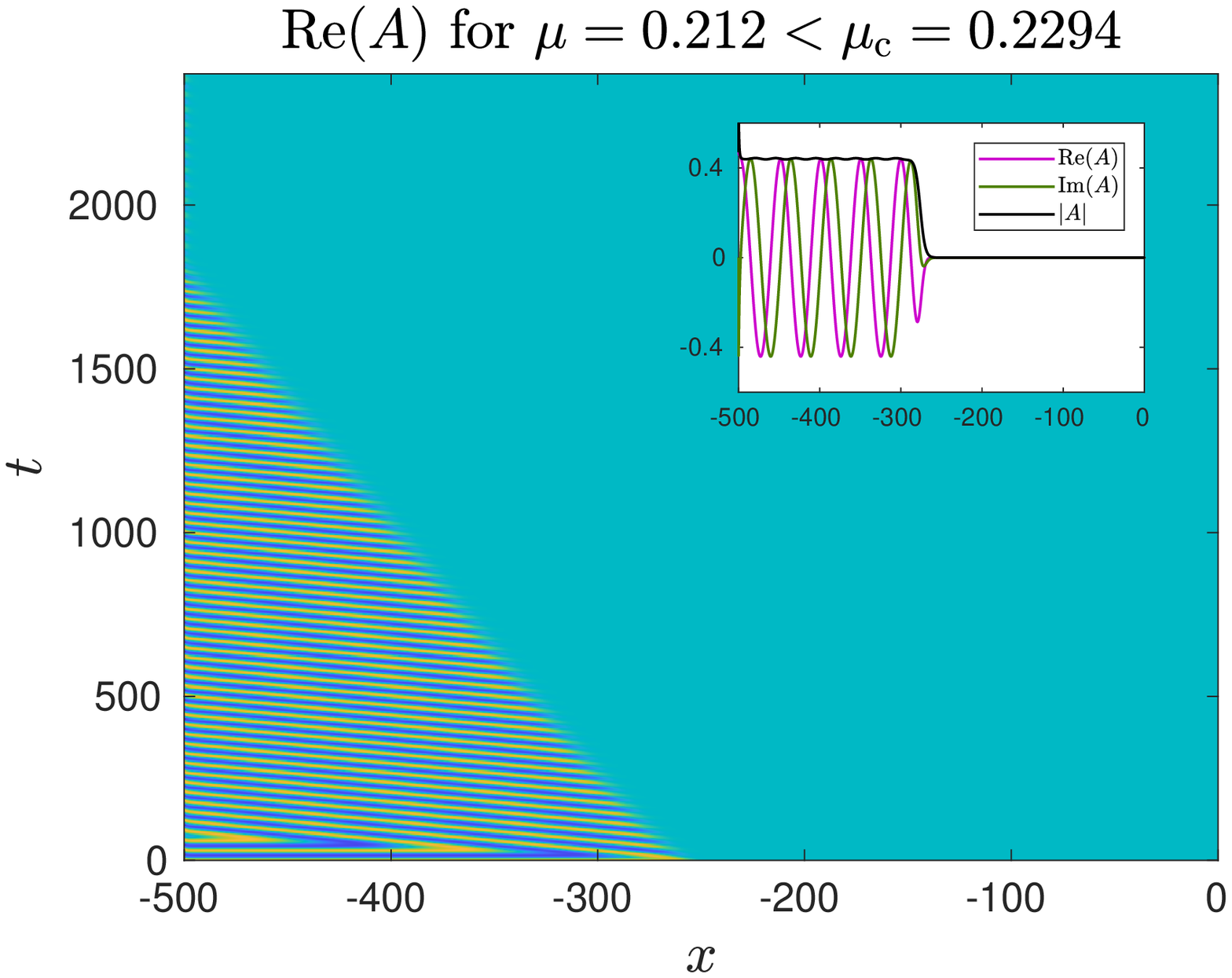}\hfill
 \includegraphics[height=2.1in]{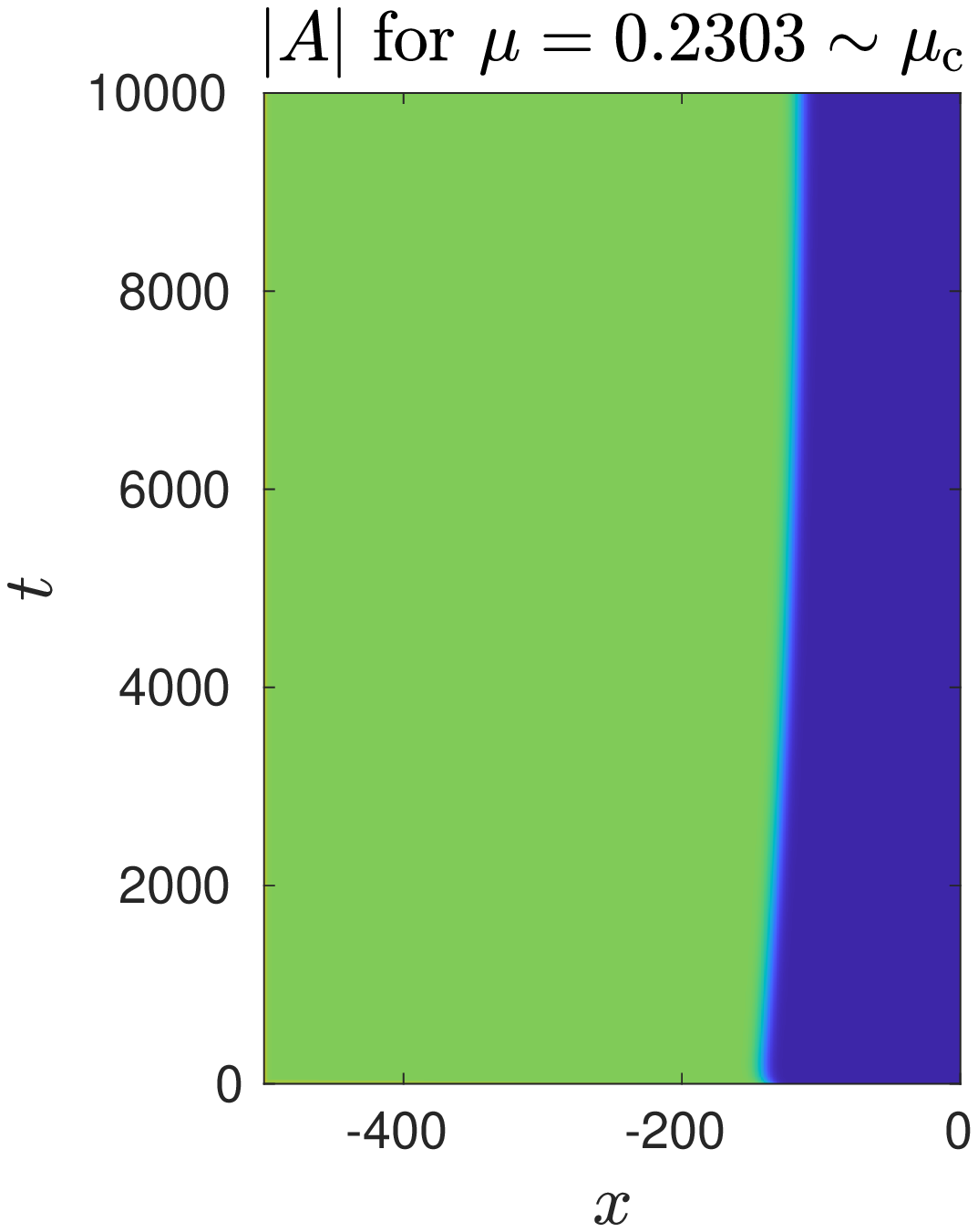}\hfill
 \includegraphics[height=2.1in]{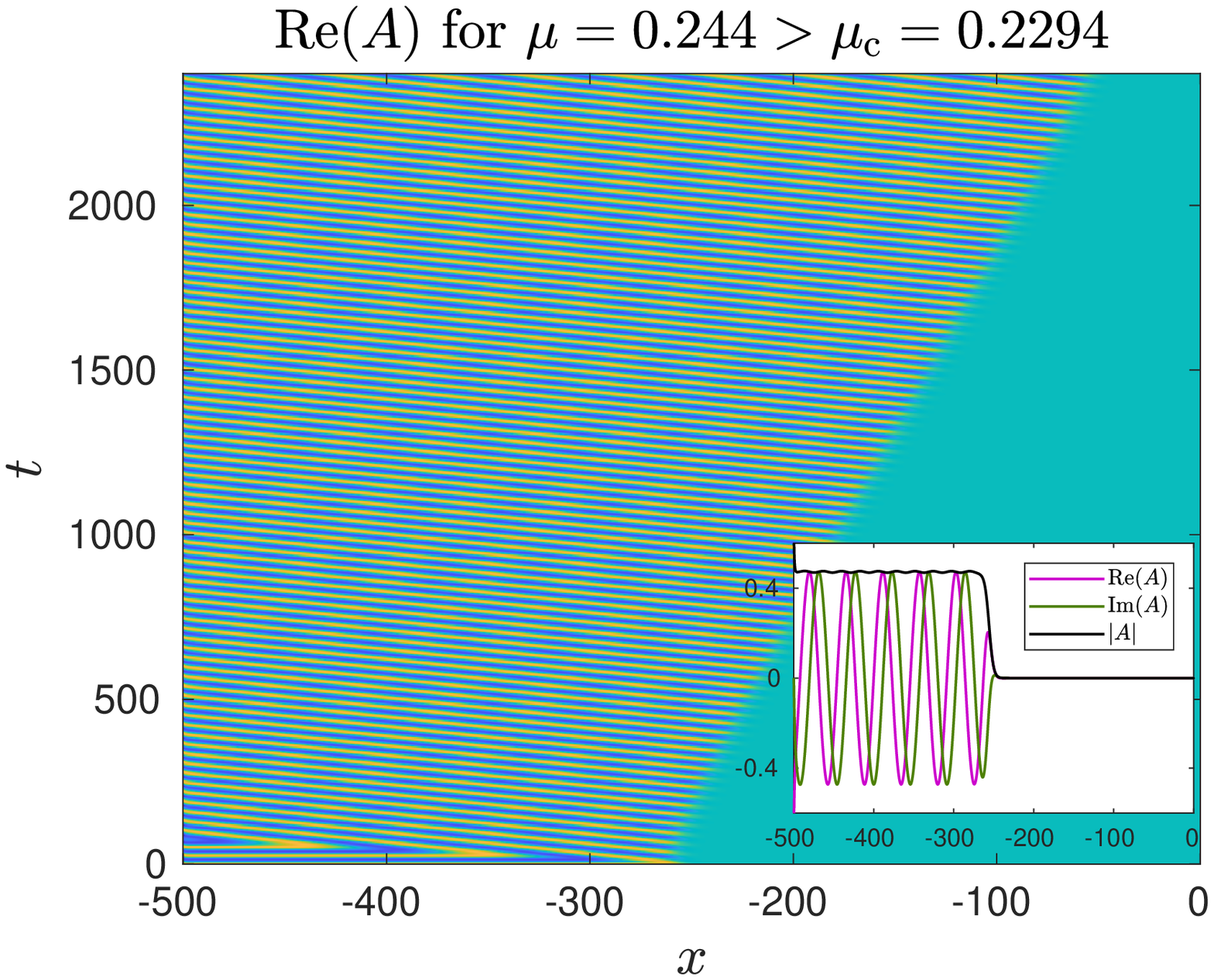}
 \caption{Space-time plots for CGL \eqref{e:cgl}, $\alpha=0.3,\,\gamma=0.3,\,d=1,\,R_0=.1,\,k_0=0.5$, with $\mu$ below (left, $\Re(A)$), close to (center, $|A|$), and above (right, $\Re(A)$) the critical value. Left and right figures show front positions diverging to left and right border,  $\Xi(\mu)=0,1$ respectively, and $\Xi(\mu)\sim 0.8$ converges in the middle figure (note the different time scale).}\label{f:ill}
\end{figure}

\textbf{Large bounded domains.}
We focus on unidirectional transport, modeled by a simple advection term, in essentially one-dimensional systems, in large bounded domains. We assume that the upstream boundary condition, say at $x=0$ suppresses instability, and that the downstream boundary condition at $x=-L$ is compatible with a nontrivial finite-amplitude state. Relating to the previous discussion, we are interested in the situation where a pulled front, with speed determined by either branched or unbranched resonances, mediates the instability in an unbounded domain. Ignoring the effects of the boundary when the front location is sufficiently far from the boundary, we then expect a \emph{hard} transition: the instability is swept out of the system when the front recedes, that is, for negative linear speeds, and the instability invades the entire domain when the linear speed is positive. One expects interaction with the boundary to be weak so that front motion is arrested only at a finite $L$-independent distance to the boundary; see Fig. \ref{f:ill} for an illustration. Such hard transitions  and steep bifurcation branches have been both observed and computed numerically, with a partial list including Langmuir-Blodgett transfer
\cite{Koepf_2014}, convection patterns with through flow \cite{PhysRevA.45.3714}, Couette-Taylor flows with axial through flow \cite{PhysRevE.53.4764}; see also \cite{knobloch}, \cite{chomaz}, and \cite{vansaarloos03} for further references. 

A standard bifurcation analysis in large domains, using for instance Lyapunov-Schmidt or center-manifold reduction to a kernel of the linearized operator, is hindered by the fact that spectra of the linearization at a trivial state form clusters that accumulate on absolute spectra \cite{ssabs} as $L\to\infty$, which in our case of unidirectional transport are strictly to the left of the continuous spectrum in the complex plane. In-between continuous and absolute spectra lies a region of pseudo spectrum that indicates sensitivity of the system to small disturbances; see for instance \cite{knobloch,ssbasin} for an analysis of the dynamics near such transitions. Moreover, such a perturbative analysis will find solutions that are small in amplitude, while the saturated nonlinear state is typically of order one at the onset of absolute instability. The bifurcation analysis that we pursue here therefore describes the nonlinear bifurcating state using the nonlinear front as an ingredient that is matched with the boundary conditions. In the case of branched resonances, such matching, only for the upstream boundary, leads to nonlinear global modes  \cite{cc96,cc97,gs1}, which, near onset, resemble the nonlinear invasion front arrested at a finite distance from the upstream boundary.

%
%

\textbf{Main objective.}
We wish to investigate instabilities in large bounded domains in the case of pulled front invasion for both branched and unbranched resonances, hoping to predict  universal features of bifurcation diagrams in large domains for branched and unbranched resonances.
We therefore introduce a class of model problems, next, and then state our main results informally in terms of a solution measure that tracks the portion of the domain occupied by the nontrivial state.

\textbf{Model problems.}
The simplest model of a transition from convective to absolute instability is the scalar amplitude equation, usually referred to as Allen-Cahn, real Ginzburg-Landau, or (generalized) Fisher-Kolmogorov-Petrovsky-Piscounov equation (KPP), 
\begin{equation}\label{e:kpp}
  u_t=u_{xx}+u_x + \mu u - u^3, \quad -L<x<0.
\end{equation}
We choose Dirichlet boundary conditions, suppressing an instability upstream and modeling a downstream disturbance,
\begin{equation}\label{e:kppbc}
u=u_0,\quad x=-L,\qquad \qquad u=0,\quad x=0.
\end{equation}
We will discuss later how results are largely independent of boundary conditions and in particular the size of $u_0\neq 0$. 

More realistic models for instabilities in fluid flows assume oscillatory instabilities captured by a complex amplitude solving the complex Ginzburg-Landau equation,
\begin{equation}\label{e:cgl}
  A_t=(1+\rmi\alpha)A_{xx}+A_x + (\mu+\rmi\beta) A - (1+\rmi\gamma)A|A|^2, \quad -L<x<0.
\end{equation}
which we equip similarly with Dirichlet or  gauge-invariant boundary conditions:
\begin{align}
A=A_0, \quad x=-L,\qquad \qquad &A=0,\quad x=0, \quad \text{or}\label{e:cglbc1}\\
|A|^2_x=R_0, \ \mathrm{arg}(A)_x=k_0,\quad x=-L,\qquad \qquad &A=0,\quad x=0. \label{e:cglbc2}
\end{align}
In both cases, the transition from convective to absolute instability is caused by a branched resonance, that is,  a pinched double root that crosses the imaginary axis at $\mu_\mathrm{br}=\frac{1}{4}$, \eqref{e:kpp}, and $\mu_\mathrm{br}=\frac{1}{4(1+\alpha^2)}$, \eqref{e:cgl}. 

The somewhat peculiar choice of boundary conditions in \eqref{e:cglbc2} simplifies numerical continuation since the boundary conditions preserve the gauge invariance $A\mapsto \rme^{\rmi\varphi}A$. As a consequence, one can study the invasion process by investigating solutions of the form $A(t,x)=A_*(x)\rme^{\rmi\omega t}$, which in turn solve an ODE boundary value problem. In particular, in this case the parameter $\beta$ is redundant and can be set to zero. 

Unbranched resonances typically arise when more than one linear mode is unstable or only weakly stable. Such a situation is conveniently captured through coupled amplitude equations, although, as mentioned in the introduction, we do not claim that the form of the equation gives a universal description of the convective-to-absolute transition.  Our main motivating example then is the system 
\begin{alignat}{2}
  u_t=&\  u_{xx} +u_x + \mu u - u^3 + |A|^2, \quad &-L<x<0,\nonumber\\
  A_t=&\ d(1+\rmi\alpha)A_{xx}+A_x + (\rho+\rmi\beta) A - (1+\rmi\gamma)A|A|^2, \quad &-L<x<0.\label{e:kppcgl}
\end{alignat}
with boundary conditions 
\begin{equation}\label{e:kppcglbc}
|A|^2_x=R_0,\ \mathrm{arg}(A)_x=k_0,\ u=u_0,\quad x=-L,\qquad \qquad A=0,\ u=0,\quad x=0.
\end{equation}
We will also study boundary conditions that break the gauge symmetry in $A$, \eqref{e:cglbc1}, and more generic coupling terms.

To further an analytical understanding, we also consider the case where $\alpha=\gamma=\omega=0$, and $A\in\R$, which reduces to a system of coupled scalar amplitude equations,
\begin{align}\label{e:kppkpp}
  u_t=&\ u_{xx}+u_x + \mu u - u^3 + v^\kappa, \quad -L<x<0,\nonumber \\
  v_t=&\ d v_{xx}+v_x + \rho v - v^3, \quad -L<x<0.
\end{align}
with boundary conditions 
\begin{equation}\label{e:kppkppbc}
v=v_0,\ u=u_0,\quad x=-L,\qquad \qquad v=0,\ u=0,\quad x=0.
\end{equation}
The exponent $\kappa$ determines the type of resonance that is relevant: clearly, $\kappa=2$ -- and  therefore 2:1-resonances -- would be obtained from \eqref{e:kppcgl}, but we shall also study $\kappa=1$, which leads to 1:1-unbranched resonances. 

Lastly, motivated by the exposition in \cite{holzerscheel14}, we also study the simplest model problem that produces unbranched resonant instabilities, 
\begin{alignat}{2}\label{e:cpw}
  u_t=&\  (1-\mu)u_x +  u - u^3 + v, \quad &-L<x<0,\nonumber \\
  v_t=&\  - v_x -  v, \quad &-L<x<0.
\end{alignat}
with $\mu<1$ and boundary conditions 
\begin{equation}\label{e:cpwbc}
v=v_0,\quad x=-L,\qquad \qquad u=0,\quad x=0.
\end{equation}
%

 
\textbf{Main result --- informal.} We measure amplitudes through the support of the solution, or, more precisely, the region where the amplitude of the solution exceeds some fixed, small threshold,  within the interval $[-L,0]$. Assuming predominantly leftward transport, the support of the solution grows as the control parameter $\mu$ increases past an instability threshold $\mu_\mathrm{c}$, determined by the fact that the speed of an invasion front in the unbounded domain changes sign, that is, the front invades the domain for $\mu>\mu_\mathrm{c}$. Writing $X(\mu)$ for the position of the interface, that is, the right boundary of the region where $|u(x)|>\delta$ for some small, positive $\delta$, we define  $\Xi(\mu)=(X(\mu)+L)/L$, the fraction of the domain occupied by the nontrivial state,  as our solution measure. 

\begin{figure}
\centering
\includegraphics[height=0.6in]{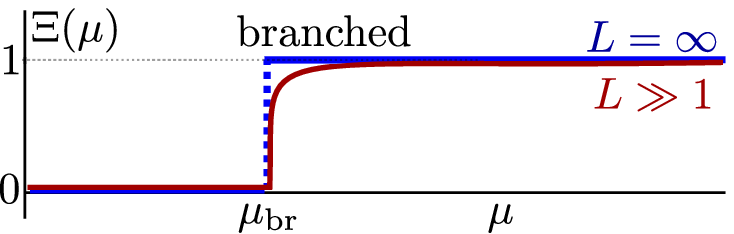}
\hspace*{0.8in}
\includegraphics[height=0.6in]{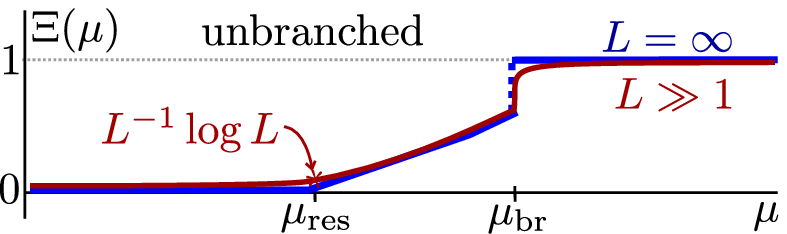}
\caption{Typical bifurcation diagrams for branched (left) and unbranched (right) resonances, as described in our main results. }\label{f:bifschem}
\end{figure}

A brief summary of our main results is shown in Fig. \ref{f:bifschem} and reads as follows. 
\begin{itemize}
 \item \emph{branched resonance, \eqref{e:kpp} or  \eqref{e:cgl}:} 
 \[\Xi(\mu)\sim \left\{
 \begin{array}{ll} 
    0,\ &\mu<\mu_\mathrm{br}\\
    1,\ &\mu>\mu_\mathrm{br},
 \end{array}\right.
\]
in the limit $L\to\infty$. For any finite $L$,  $\Xi(\mu)$ is smooth  with an increase from $\eps>0$, small, to $1-\eps$ on a parameter interval of width $\rmO(L^{-1/2})$ as $L \to \infty$.
\item \emph{unbranched resonance, \eqref{e:kppcgl}, \eqref{e:kppkpp}, or \eqref{e:cpw}:} 
\[
 \Xi(\mu)=\frac{\nu_0^\mathrm{u}(\mu)}{\nu_0^\mathrm{u}(\mu)+\nu_1^\mathrm{s}(\mu)},
\]
in the limit $L\to\infty$, a smooth function for $\mu\geq \mu_\mathrm{res}$. Here, $\nu_0^\mathrm{u}(\mu)$ characterizes the distance of the spatial modes from resonance and $\nu_1^\mathrm{s}(\mu)$ is a decay rate of boundary layers.    For finite $L$, $\Xi(\mu_\mathrm{res})=\rmO(L^{-1}\log L)$.
\end{itemize}

\textbf{Technical contribution.} We find bifurcation diagrams of stationary solutions using a spatial dynamics technique. We construct solutions in the bounded domain by matching the nonlinear pulled front with the downstream boundary via a boundary layer and, crucially, with the upstream boundary after a geometric desingularization procedure which identifies branched resonances as a saddle-node bifurcation and unbranched resonances as a transcritical bifurcation in projective space. In suitable coordinates, the construction of solutions that satisfy boundary conditions then reduces to a heteroclinic gluing analysis. Our analysis is novel and rigorous in the simple model examples, although formal expansions in the unbranched case appear to be known in specific situations; see e.g. \cite{knobloch}. It also isolates conceptual assumptions which guarantee that in other, possibly more complex situations analogous statements hold, including in particular a large class of boundary conditions and systems posed on long cylindrical domains. We complement the analysis with a numerical continuation analysis that confirms our predictions in simple models and, somewhat, beyond. 

\textbf{Outline.} We demonstrate the hard onset of instability in the case of branched resonances,~\S\ref{s:b}, and turn to unbranched resonances in \S\ref{s:u}. We conclude with a discussion in \S\ref{s:d}. 

\textbf{Acknowledgments.} C.D.,  A.Sm., and A.Sc. were supported through grant NSF DMS-1907391. M. A. was supported through  the NSF GRFP, Award 00074041.

\section{Branched resonances --- hard onset}\label{s:b}

The transition from convective to absolute instability in these model problems with a branched resonance is fairly well studied and various asymptotics have been previously derived in the literature; see in particular \cite{knobloch}. We are however not aware of a somewhat rigorous derivation of the $\Xi$-asymptotics, particularly in the generic setting with the presence of boundary layers. The analysis here is also closely mimicked in the next section on unbranched resonances.

\subsection{Analysis and predictions}\label{s:ap}

We first analyze \eqref{e:kpp} in unbounded and large bounded domains, before turning to \eqref{e:cgl}. 
We start with the linear KPP equation and its dispersion relation for solutions $\exp(\lambda t + \nu x)$,
\[
u_t=u_{xx}+u_x+\mu u,\quad x\in\R, \qquad\qquad D(\lambda,\nu)=\nu^2+\nu +\mu-\lambda.
\]
Solutions  with initial data $u_0(x)$ are given through the explicit ``heat kernel'', which in turn can be constructed through inverse Laplace transform of the resolvent,
\[
 u(t,x)=\frac{1}{2\pi\rmi}\int_{\lambda \in \Gamma} \int_y \rme^{\lambda t} G_\lambda(x-y) u_0(y)\rmd y\rmd\lambda,
\]
where $\Gamma$ is a curve in the complex plane to the right of the spectrum of $\mathcal{L}=\partial_{xx}+\partial_x+\mu$ with sectorial asymptotics $r\rme^{\pm\rmi\theta}$, $\theta\in (\pi/2,\pi)$, and $(\mathcal{L}-\lambda)G_\lambda(x)=-\delta(x)$. The exponential growth rate of $u(t,x)$ is bounded by the maximal real part of $\Gamma$ and we therefore deform the contour to the left in $\C$ as far as possible. With a view on restricting to a compact interval, later, we think of compactly supported initial conditions and observations $x$ in a bounded region, so that we may deform contours until we reach singularities of the \emph{pointwise} Green's function $G_\lambda(\xi)$. In this simple case, 
\[
G_\lambda(\xi)=\frac{1}{\nu_+-\nu_-} \rme^{\nu_\pm\xi}, \quad\text{ for } \mp\xi>0 ,
\]
where $\nu_\pm=\nu_\pm(\lambda)$ are roots to $D(\lambda,\nu)=0$ with $\pm\Re\nu_\pm>0$ for $\lambda\gg 1$. 
Clearly, $G_\lambda$ is analytic up to $\lambda_\mathrm{dr}=\mu-1/4$, the pinched double root, where $\nu_+=\nu_-$

More generally and more directly, one looks for pinched double roots by solving 
\[
 D(\lambda,\nu)=0,\qquad \partial_\nu D(\lambda,\nu)=0, 
\]
with a pinching condition $\nu_\pm(\lambda)\to\pm\infty$ for $\Re\lambda\to\infty$; see \cite{bers84,holzerscheel14} for more details and references. 

Continuing the analysis in the linearized KPP setting, we find $(\lambda_\mathrm{dr},\nu_\mathrm{dr})=(\mu-\frac{1}{4},-\frac{1}{2})$ such that we have pointwise stability of the origin for $\mu\leq \mu_\mathrm{br}=\frac{1}{4}$. 

In this specific case, one readily finds explicitly, by solving the Sturm-Liouville eigenvalue problem,  that homogeneous Dirichlet boundary conditions at $x=-L$ and $x=0$ suppress the instability slightly for large $L$, with a corrected onset $\tilde{\mu}_\mathrm{br}=\frac{1}{4}+\frac{\pi^2}{L^2}$ that one can readily compute 

Rather than pursuing a bifurcation analysis from a trivial solution, which usually has limited validity and predictive strength in large domains, we pursue a conceptual analysis based on spatial dynamics that also generalizes to systems and resembles the point of view taken in the case of unbranched resonances. We look for solutions as orbits in the phase plane connecting subspaces determined by the boundary conditions. 

\begin{figure}
 \includegraphics[width=0.3\textwidth]{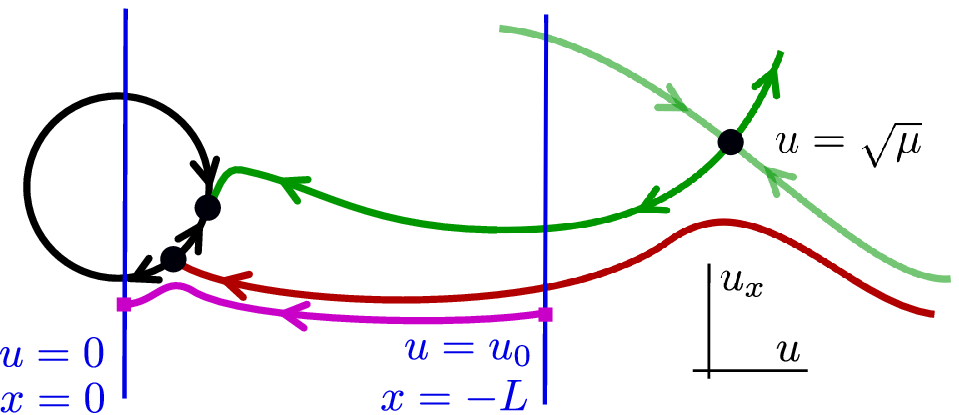}\hfill
 \includegraphics[width=0.3\textwidth]{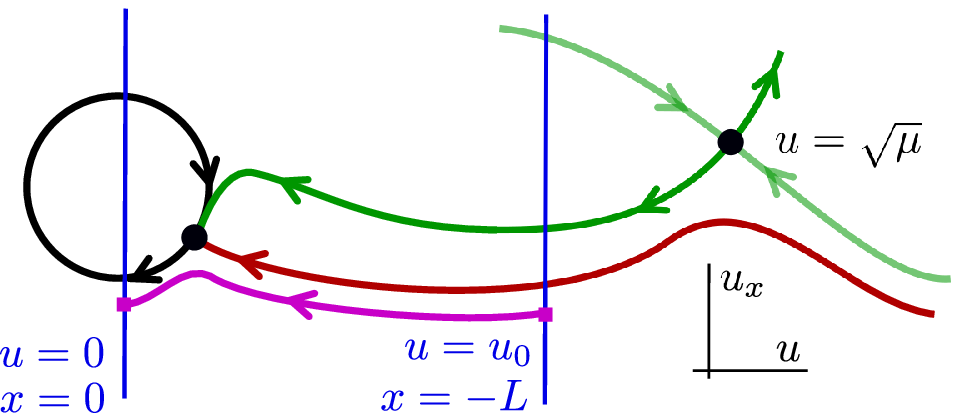}\hfill
 \includegraphics[width=0.3\textwidth]{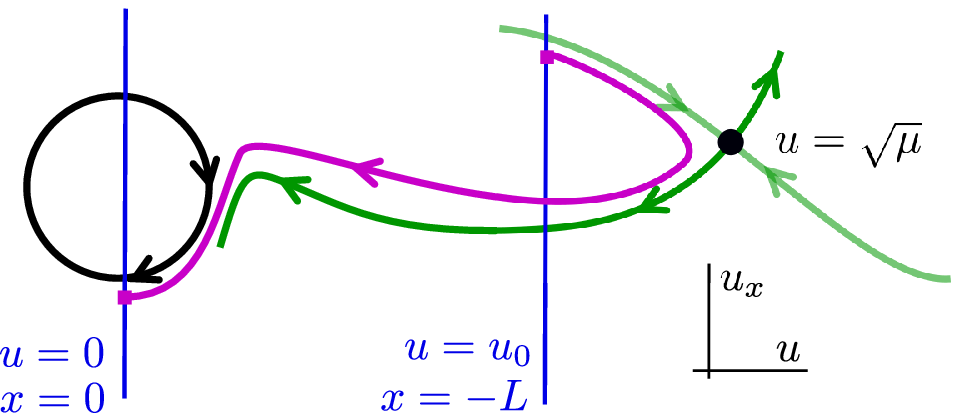}
 \caption{Phase plane for \eqref{e:kppstst} at $\mu\lesssim1/4$, $\mu=1/4$, and $\mu\gtrsim 1/4$ (left to right). Solutions (magenta) connecting boundary conditions (blue) with long  flight times $L$. The flight time near the blown up equilibrium $u=0$ decreases with $(\mu-1/4)^{-1/2}$, and additional flight time is added through a long passage near the equilibrium $u=1$. }\label{f:kppphpl}
\end{figure}
\begin{figure}[h!]
 \centering\includegraphics[width=0.2\textwidth]{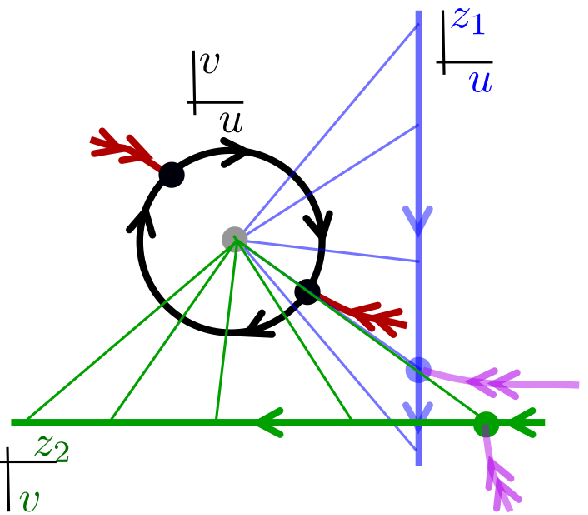}\hspace*{0.1\textwidth}
  \raisebox{0.2in}{\includegraphics[width=0.32\textwidth]{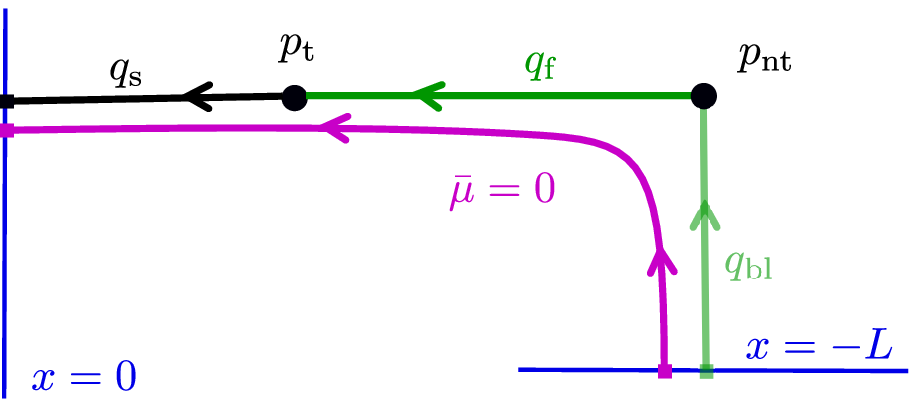}}
  \caption{Visualization of the coordinate changes from polar coordinates to projective coordinates (left): Blown up equilibrium (black) with singular eigenspaces and strong stable manifolds (red). Projection onto $(u,z_1)$-coordinates (blue) and $(v,z_2)$-coordinates. Also shown a schematic of the chain of heteroclinic orbits (right) as detailed in the itemized list of building blocks in the text. }\label{f:proj}
\end{figure}

We therefore consider the steady-state equation
\begin{align}
u_x=&\,v\nonumber\\
v_x=&\,-v-\mu u + u^3.\label{e:kppstst}
\end{align}
At $\mu=1/4$, the equilibrium $u=v=0$ possesses a double eigenvalue $\nu=-1/2$ which we desingularize using polar coordinates $(u,v)=r(\cos\varphi,\sin\varphi)$, $(r,\varphi)\in [0,\infty)\times S^1$. The origin $u=v=0$ becomes an invariant circle $\{0\}\times S^1$, with equilibria given by the eigenspaces of the linearization. The corresponding saddle-node bifurcation at $\mu=1/4$ leads to slow periodic orbits on the circle for $\mu>1/4$. The resulting phase portraits (blowing up the origin in polar coordinates $(r+1)(\cos(\varphi),\sin(\varphi))$ is plotted in the coordinate plane in Fig.~\ref{f:kppphpl}. Instead of working with angles and radii in polar coordinates, it is algebraically convenient to use stereographic projection charts 
$(u,z_1)$, with $z_1=v/u$, and $(v,z_2)$, with $z_2=u/v$. The origin, blown up to a sphere in polar coordinates, is now represented by the lines $\{0\}\times\R$. The equations take the explicit form
\begin{equation}
\left\{ \begin{array}{rl}
         u'=&uz_1\\
         z_1'=&-\left(z_1+\frac{1}{2}\right)^2-\left(\mu-\frac{1}{4}\right)+u^2,
        \end{array}\right. \qquad\qquad 
\left\{ \begin{array}{rl}
         v'=&v\left(-1-\mu z_2 + v^2z_2^3\right)\\
         z_2'=& 
         \left(\frac{z_2}{2}+1\right)^2+\left(\mu-\frac{1}{4}\right)z_2^2-v^2z_2^4;
        \end{array}\right. \label{e:proj}
\end{equation}
see Fig.~\ref{f:proj} for an illustration. Key ingredients to the bifurcation, schematically illustrated in Figure~\ref{f:proj}, are:
\begin{enumerate}
 \item the trivial equilibrium $p_\mathrm{t}$, shown as the invariant circle $r=0$;
 \item a saddle-node bifurcation of equilibria on the critical circle reflecting the branched resonance in a collision of eigenspaces;
 \item the nontrivial equilibrium $p_\mathrm{nt}$, hyperbolic in this spatial dynamics formulation since stable in temporal dynamics;
 \item a heteroclinic orbit $q_\mathrm{f}$, the invasion front on the unbounded domain, between nontrivial equilibrium and trivial equilibrium, \emph{not} contained in the strong stable manifold of the saddle-node equilibrium at criticality;
 \item a boundary subspace for $x=0$ that does \emph{not} contain the saddle-node equilibrium at criticality; 
 \item a ``singular heteroclinic'' $q_\mathrm{s}$ from the saddle node to the boundary condition in the invariant circle;
 \item a boundary subspace for $x=-L$ that intersects the stable manifold of the nontrivial equilibrium transversely at criticality yielding a ``boundary layer heteroclinic'' $q_\mathrm{bl}$. 
\end{enumerate}
Finding these objects is explicitly possible in the KPP equation but can be readily verified computationally in more complicated systems. We remark that the fact in (iv) that $q_\mathrm{f}$ is not contained in the strong stable manifold of $p_\mathrm{t}$ is equivalent to requiring the absence of a pole of the resolvent for the linearization at a critical pulled front. Such a pole, reflected in a zero of an Evans function would usually indicate a transition from pulled to pushed front propagation. In the KPP case, it is equivalent to the fact that the front possesses asymptotics $(ax+b)\rme^{-x/2}$ with $a\neq 0$. Also note that existence of a boundary layer (vii) can be guaranteed for all values of $u_0$, in particular for homogeneous Dirichlet boundary conditions, as well as for Neumann or Robin boundary conditions, implying that the asymptotics derived below are universal. Of course the fine structure of the bifurcation at small amplitude, for instance if and how the pitchfork symmetry is broken, depends on the precise form of the boundary conditions. 

With these assumptions, one finds orbits that follow the chain of ``heteroclinics'', from boundary condition at $x=-L$ to the nontrivial equilibrium, then from the non-trivial equilibrium to the trivial equilibrium, and ultimately from the saddle-node equilibrium to the boundary subspace. Those trajectories spend long times near the two equilibria included in the heteroclinic and times $\rmO(1)$ in between the equilibria. Passage times near the saddle  node  are readily found as $L_0=\pi (\mu-1/4)^{-1/2}+\rmO(1)$. Solutions with flight times $L$ between the boundary conditions therefore need to add a plateau near the nontrivial equilibrium $u=\sqrt{\mu}$ of length $L-\pi (\mu-1/4)^{-1/2}+\rmO(1)$, which gives the prediction
\begin{equation}\label{e:kpppredict}
 \Xi(\mu)=1-\frac{\pi}{L\sqrt{\mu-\mu_\mathrm{br}}}.
\end{equation}
As a result, we find 
\begin{equation}
\label{e:kppthy}
\Xi(\mu)\sim \mathrm{max}\,\left(0,1-\frac{\pi}{L\sqrt{\mu-\mu_\mathrm{br}}}\right), \text{ for } \mu\gtrsim \mu_\mathrm{br}. 
\end{equation}
which converges to $1$ for $\mu>\mu_\mathrm{br}$, fixed, as $L\to\infty$. 
In a more general setting, for a branched double root in the dispersion relation at $\lambda = \bar{\mu}\sim 0$, with expansion of the dispersion relation $D(\lambda,\nu)=\lambda -d(\nu-\nu_\mathrm{br}^2)^2+\bar{\mu}$, the saddle-node bifurcation unfolds as $dz_1'=-d(z_1-\nu_\mathrm{br})^2-\bar{\mu}$ and we predict that 
\begin{equation}
 \Xi(\mu)=\mathrm{max}\,\left(0,
1-\frac{\pi}{L\sqrt{\bar{\mu}/d}}\right).
\end{equation}


We believe that the ingredients to this analysis are quite generally satisfied near onset of instability and the predictions are therefore universally valid across a wide range of systems; see for instance \cite[Thm. 2]{avery2020universal}, where robustness of pulled fronts was established. 

For the complex Ginzburg-Landau equation, the problem is very similar. The key difference is that the pinched double root is complex $\Im\lambda_\mathrm{dr}\neq 0$, $\Im\nu_\mathrm{dr}\neq 0$. As a consequence, the resonant subspace is 4-dimensional and the desingularization leads to dynamics on $[0,\infty)\times S^3$. The gauge symmetry acts however on $S^3$ via the Hopf fibration and the problem reduces to $[0,\infty)\times S^2\sim [0,\infty)\times \bar{\C}$, with $\bar{\C}$ the Riemann sphere, with singular heteroclinics given by solutions to $z'=-z^2\in\bar{\C}$. An equivalent analysis was carried out in  \cite[\S3]{gs1}  which we refer to for more details of the choice of coordinates.

Since the reduced equation is simply the complexification of the real saddle-node, we find the same leading order term in passage times, 
\begin{equation}
\label{e:cglthy}
\Xi(\mu)\sim \mathrm{max}\,\left(0,1-\frac{\pi}{L\sqrt{\mu-\mu_\mathrm{br}}}\right), \text{ for } \mu\gtrsim \mu_\mathrm{br},
\end{equation}
where this time $\mu_\mathrm{br}=\frac{1}{4(1+\alpha^2)}$; see also \cite[\S2.3]{knobloch}. 

While passage times near the (complex) saddle-node can be computed, other parts of the heteroclinic bifurcation diagram are not as easily established rigorously. Existence of $q_\mathrm{f}$ is not generally known outside of a regime where $\alpha,\gamma\sim 0$, although the existence problem reduces to a shooting problem in a 3-dimensional ODE with good evidence for existence across all parameters. There also does not appear to be evidence for existence of pushed fronts. Absent instabilities of the primary front, the main limitation to universal validity of our expansion then are boundary layers. Since the pulled fronts in CGL select wave trains with group velocities pointing away from the front interface \cite{vansaarloos03,gs1}, boundary layers $q_\mathrm{bl}$ are merely boundary sinks and are expected to exist in a robust fashion for a family of wavetrains and frequencies \cite{ssdefect}, in particular the ones selected by the invasion front. 

Lastly, we note that in both KPP and in CGL, there are multiple singular heteroclinic orbits, corresponding to multiple (half-)rounds on the singular circle (sphere). The associated profiles are typically unstable but accessible to numerical continuation. Scalings of positions are similar, with passage times near the saddle-node replaced simply by multiples, leading to 
\begin{equation}
\label{e:cglthyj}
\Xi(\mu)\sim \mathrm{max}\,\left(0,1-j\frac{\pi}{L\sqrt{\mu-\mu_\mathrm{br}}}\right), \text{ for } \mu\gtrsim \mu_\mathrm{br}, \qquad j=1,2,3,\ldots
\end{equation}

\subsection{Computational bifurcation analysis}
\begin{figure}[h!]
\centering \includegraphics[width=0.3\textwidth]{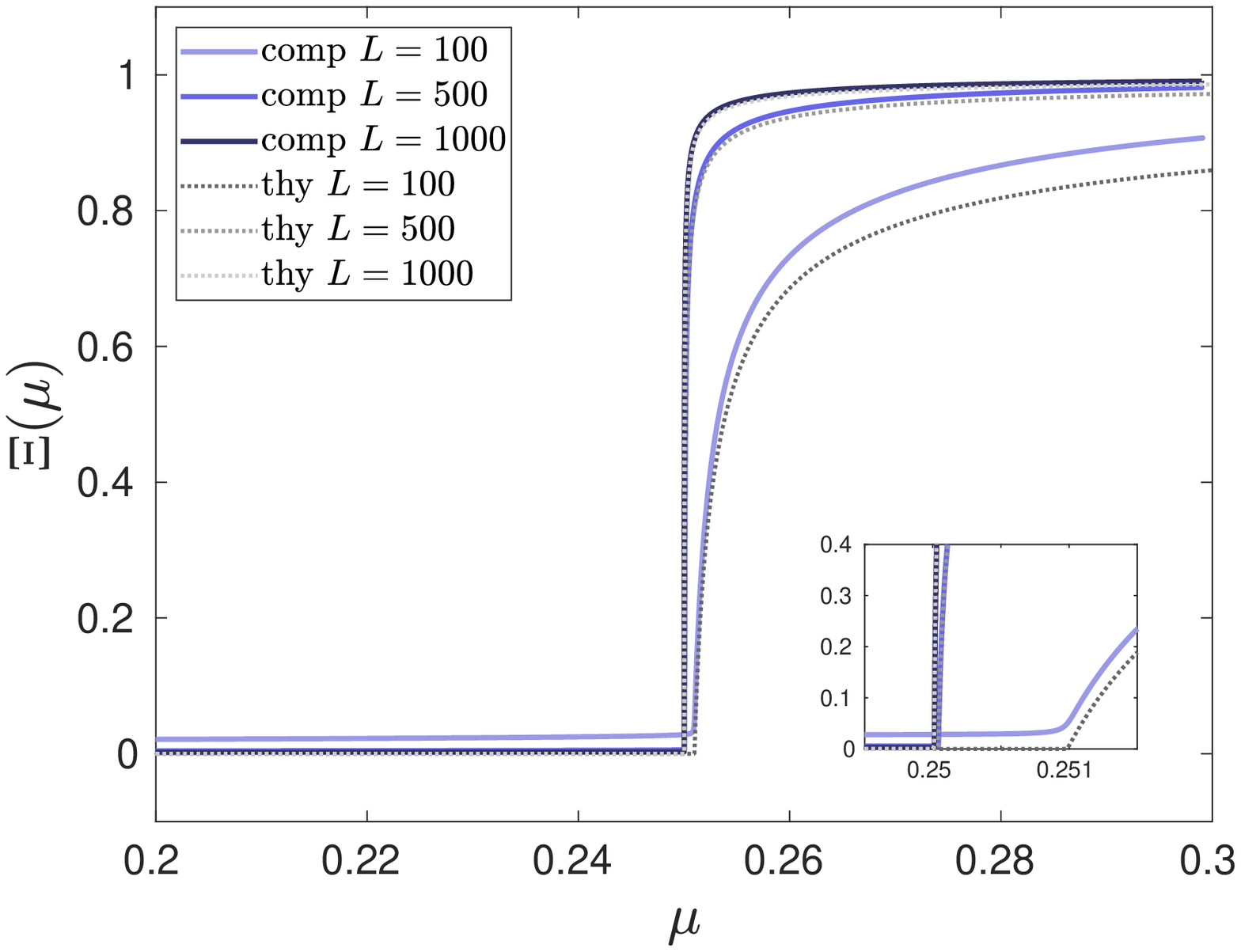}\hfill \includegraphics[width=0.3\textwidth]{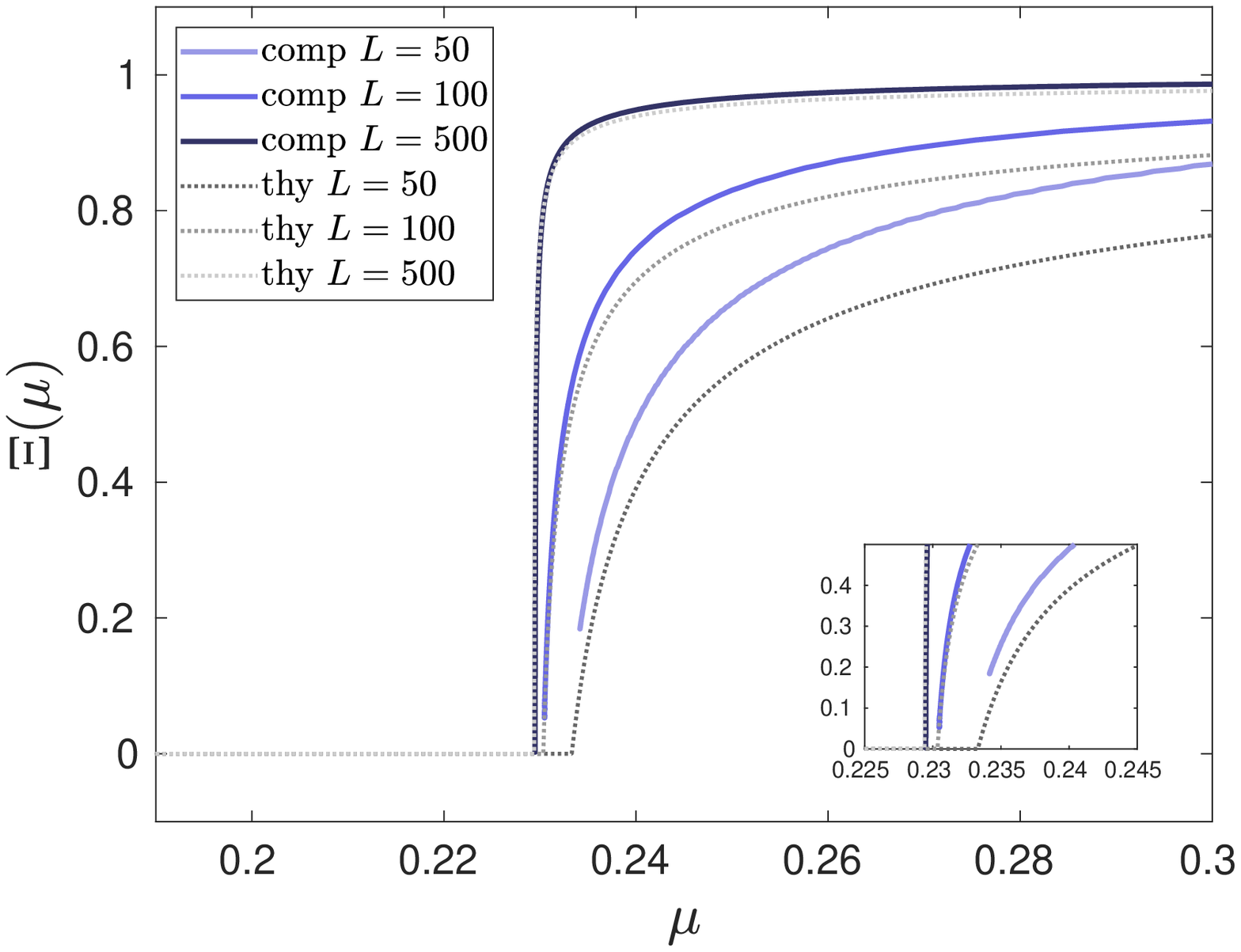}\hfill \includegraphics[width=0.3\textwidth]{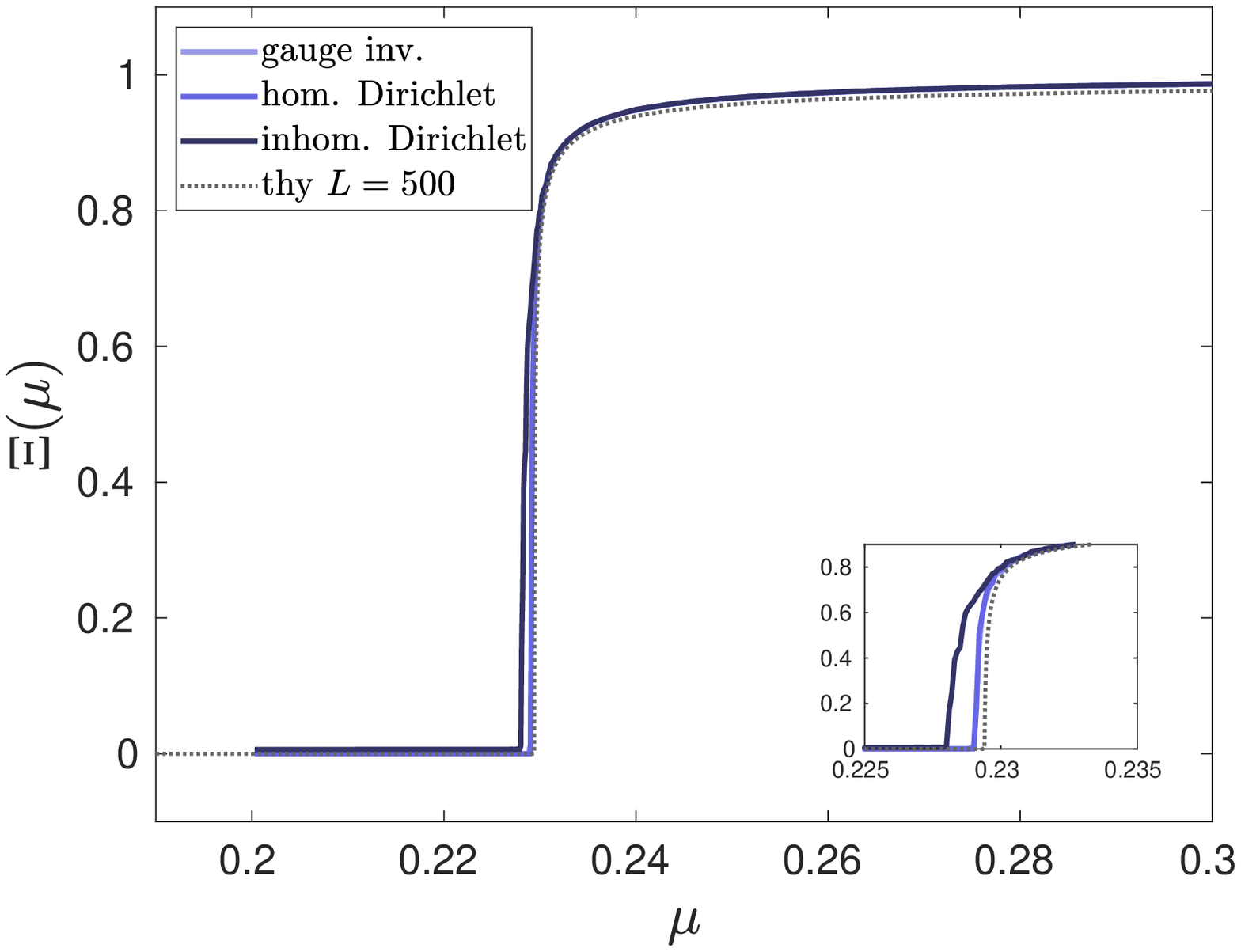}
 \caption{Numerical bifurcation diagrams for KPP (left, \eqref{e:kpp}) and CGL (center, \eqref{e:cgl}), and comparison with theory \eqref{e:kppthy} and \eqref{e:cglthy}, respectively, showing excellent agreement for the front position. Parameters for CGL are $\alpha=0.3,\,\gamma=0.3,\ R_0=0.1,\,k_0=-0.5$. Bifurcation diagrams from direct simulations of CGL (right) with gauge-invariant boundary conditions \eqref{e:cglbc2} as in center figure, homogeneous and inhomogeneous Dirichlet boundary conditions ($R_0=0,\,R_0=1$, \eqref{e:cglbc1}).}\label{f:kppcglcont}
\end{figure}

We discretized the boundary-value problems using second order finite differences with $h=10^{-2}$ and continued solution branches using secant continuation. Results are shown in Fig.~\ref{f:kppcglcont}. Figures include only the positive branch in the KPP equation and the branch with $j=1$ in CGL \eqref{e:cglthyj}. The branch with plateau near $-\sqrt{\mu}$ for $\mu>\mu_\mathrm{br}$ undergoes a saddle-node near $\mu=1/4$ upon decreasing $\mu$  to an unstable branch. For the unstable branch, $\Xi(\mu)$ is significantly smaller. We also studied CGL \eqref{e:cgl} with the same gauge-invariant boundary conditions \eqref{e:cglbc2} in direct simulations, comparing in particular with homogeneous Dirichlet boundary conditions and inhomogeneous Dirichlet boundary conditions. In the latter case, the amplitude $|A|$ is time-periodic and the interface location is oscillating in time, albeit with an extremely small amplitude. The graphs shown are instantaneous measurements in time. Since front dynamics are extremely slow, we continued dynamically, that is, we let the front position relax for a fixed parameter value until the measured speed was smaller than $5\cdot 10^{-4}$ and then decreased the parameter, using the last simulation data as initial condition. 
Clearly, for $u_0=0$, homogeneous Dirichlet boundary conditions, the system possesses a pitchfork symmetry. Nonzero $u_0$ breaks this symmetry leading to an imperfect pitchfork bifurcation, where the perturbation effect is very small even for finite $u_0$ due to the presence of downstream transport. Since in a pitchfork bifurcation, secant continuation typically continues the nontrivial branch as a saddle-node, very small step sizes are necessary to resolve the imperfection and continue toward the trivial branch and $\mu<\mu_\mathrm{br}$.  A second numerical difficulty is caused by the necessity to resolve the exponential tails of solutions near $x=0$, limiting the possible allowed sizes of $L$ due to the occurrence of underflow. In fact, the size of the layer $\Xi(\mu)$, as can be seen from the analysis and Fig.~\ref{f:kppphpl},  is determined by a slow passage near a fold in projective space, all for amplitudes of $(u,u_x)$ that are $\rmO(\rme^{-L/2})$, which limits domain sizes to about $L=1000$. 

In the case of CGL, the set of solutions is significantly more complex, in particular when varying other system parameters. Of course, the onset of instability in the complex Ginzburg-Landau equation will lead to dynamics as complicated as the dynamics of the Ginzburg-Landau equation, which can be difficult to describe, in particular in the regime where coherent solutions are all unstable due to the Benjamin-Feir instability. We therefore focus on small values of linear and nonlinear dispersion $\alpha$ and $\gamma$, here.  Direct numerical simulations show that typical initial conditions converge to truly periodic solutions (that is, the amplitude $|A(t,x)|$ is not stationary) with oscillations localized near the boundary for many choices of boundary conditions and system parameters, in particular for most choices of Dirichlet boundary conditions $|A|^2=R_0$ at $x=0$. We did not attempt to resolve the subtle exponentially small effect near the touchdown at $\Xi\sim 0$ of the bifurcation curves in the CGL case. We did also find solution curves with $j\neq 0$ and good agreement with theory \eqref{e:cglthyj}. 

We are not aware of a comprehensive study of bifurcations in CGL with advection for various types of boundary downstream boundary conditions and such an analysis would be far beyond the scope of this paper. We emphasize, however, that, despite a possibly complex bifurcation structure, our predictions do give throughout an overall very accurate prediction of front interface locations in the domain.

\section{Unbranched resonances --- gradual onset}\label{s:u}
A more general mechanism for speed selection was uncovered in \cite{faye17}. While the mechanism was shown there to be present in scalar equations, it is most easily illustrated in systems representing modes with different spatio-temporal frequencies. An explicit example is \eqref{e:kppcgl}, which models the interaction of stationary and oscillatory modes. We will illustrate the mechanism in \S\ref{s:mod} while reducing the model successively until an analysis is almost explicit. We derive bifurcation diagrams in large bounded domains in \S \ref{s:analysis}, relying on similar heteroclinic bifurcation constructions as shown in Fig.~\ref{f:kppphpl} in the simplest pictures, and generalizing to higher-dimensional models under assumptions on eigenvalue configurations at equilibria and genericity of heteroclinic intersections, usually corresponding to absence of additional marginal spectrum. We finally compare in \S\ref{s:comp} with numerical continuation. 

\subsection{Models --- why resonances matter}\label{s:mod}
We first motivate the resonant instability mechanism in a simple example, and then simplify models to arrive at simple models amenable to explicit analysis. 
\textbf{Resonant source terms driving pointwise instabilities.}
We return to the analysis of  the linear KPP equation from \S \ref{s:ap}, where
\[
 u(t,x)=\frac{1}{2\pi\rmi}\int_{\lambda \in \Gamma} \int_y \rme^{\lambda t} G_\lambda(x-y) u_0(y)\rmd y\rmd\lambda,
\]
and $\Gamma$ could be deformed up to $\lambda_\mathrm{dr}=-\frac{1}{4}$. The deformation of the contour requires us to be able to continue the integral into regions where $G_\lambda\not\in L^1$, exploiting that initial data is strongly localized. On the other hand, assuming initial data with a prescribed exponential rate $u_0(x)=\rme^{\nu x}$, we find solutions $\rme^{\lambda(\nu) t + \nu x}$ that travel at the linear envelope speed $-\Re\lambda(\nu)/\Re\nu$. 
Alternatively, we can consider source terms $f(t,x)$ in the equation, 
\begin{equation}\label{e:kppf}
u_t=u_{xx}+u_x+\mu u + f,\quad x\in\R, 
\end{equation}
and find solutions via space-time convolution of $f$ and the heat kernel. Forcing can induce instabilities even if the forcing itself exhibits pointwise spatial decay. For instance, $f(t,x)=\rme^{\nu x}$ with  $1\gg -\nu> 0$ would lead to instabilities for any $\mu>0$ since it induces modes with temporal growth rate $\lambda(\nu)=\nu^2+\nu +\mu \sim \mu>0$. Key here is that the forcing is ``resonant'', in this case stationary: forcing $f(t,x)=\rme^{\nu x+\rmi t}$ would \emph{not} lead to instabilities \cite{holzerremnant}. Note that the slow spatial decay is seen as slow pointwise temporal decay after slightly changing the speed of the comoving frame, such that even temporally and spatially decaying source terms can induce instabilities. 

This mechanism is most obvious if the source term stems from an explicit coupling to an oscillatory amplitude, modeled for instance by the complex Ginzburg-Landau equation
\begin{equation}\label{e:cgls}
 A_t=d(1+\rmi\alpha)A_{xx}+A_x +(\rho+\rmi\beta) A -(1+\rmi\gamma)A|A|^2, 
\end{equation}
where we think of $\rho<\frac{1}{4d(1+\alpha^2)}$, below the threshold for absolute instability, such that solutions decay pointwise. Similarly, we consider \eqref{e:kpp} with $\mu<1/4$, below the threshold of absolute instability, but assume coupling of the two equations. We are mostly interested in coupling terms that act as sources in the equation, such as terms $g_A(u)$ in the $A$-equation and $g_u(A)$ in the $u$-equation. With the discussion above, non-oscillatory terms are the most dangerous candidates for coupling, which motivates the choice of the source term in \eqref{e:kppf} as $f(t,x)=|A(t,x)|^2$ with $A$ from \eqref{e:cgls} and yields the coupled amplitude equation\eqref{e:kppcgl}. We emphasize here that we are thinking of pointwise stability in both $u$- and $A$-equations separately, which implies, due to the fact that the coupling is nonlinear, linear pointwise stability of the origin in the system. This linear pointwise stability is in fact robust under addition of small linear coupling in both $u$- and $A$-equations, using continuity of pinched double roots as established in \cite{holzerscheel14}. In this sense, the resonant instability is completely determined by the linear part but enabled by the presence of a nonlinear coupling term mediating the resonance. 

\textbf{From CGL-KPP to KPP-KPP.}
We can obtain a simplified model assuming $\alpha=\beta=\gamma=0$, which allows us to also assume $A\in\R$ and find the coupled KPP system \eqref{e:kppkpp}, albeit with quadratic coupling, $\kappa=2$. In the linear part, since we averaged out the oscillations,  $\alpha=\beta=0$, there now also is a strong linear, 1:1-resonance leading to instabilities precisely when there is a linear coupling term, $\kappa=1$. This 1:1-resonance is a double root of the dispersion relation, solution to $D(\lambda,\nu)=0, \partial_\nu D(\lambda,\nu)=0$. In fact, the dispersion relation factors due to the block-triangular structure,
\[
 D(\lambda,\nu)=D^u(\lambda,\nu)\cdot D^v(\lambda,\nu)=(\nu^2+\nu+\mu-\lambda)(d\nu^2+\nu+\rho-\lambda),
\]
and double roots can be formed through $D^u(\lambda,\nu)=D^v(\lambda,\nu)=0$, with local expansion 
\[
 D(\lambda,\nu)=\left(D^u_\lambda\cdot(\lambda-\lambda_\mathrm{dr})+D^u_\nu\cdot(\nu-\nu_\mathrm{dr})\right)\cdot 
 \left(D^v_\lambda\cdot(\lambda-\lambda_\mathrm{dr})+D^v_\nu\cdot(\nu-\nu_\mathrm{dr})\right)=\rmO\left(|\lambda-\lambda_\mathrm{dr}|^2+|\nu-\nu_\mathrm{dr}|^2\right).
\]
In contrast to the case of a branched double root, we typically find analytic solutions $\nu(\lambda)$ locally near the double root using either the Newton polygon, or, more directly, the fact that the two solutions $\nu_\pm(\lambda)$ can be continued separately as solutions of $D^u=0$ and $D^v=0$ by the implicit function theorem. We therefore refer to such pinched double roots as \emph{unbranched resonances}. The expansion above shows however that adding coupling terms to both $u$- and $v$-equations will introduce constant terms into the dispersion relation near $(\lambda_\mathrm{dr},\nu_\mathrm{dr})$. Solving for double roots will then yield a unique solution $\nu$ from $\partial_\nu D=0$, which in turn yields locally two solutions $\lambda$ from $D(\lambda,\nu)=0$. These solutions would then be branched resonances and produce phenomena discussed in the previous section. Since such pinched double roots cannot occur through linear coupling of KPP and Ginzburg-Landau, we preserve the analogy and also avoid these pinched double roots in this simplified setup of coupled stationary modes by restricting to unidirectional linear coupling, that is, excluding linear $u$-dependence in the $v$-equation. We refer to \cite{holzerscheel14} for a more thorough discussion of such unbranched resonances and the role of the pinching condition. 

\textbf{From KPP-KPP to CPW.}
Simplifying even further, we inspect the dispersion relation near the branch point in a canonical form, eliminating $(\lambda-\lambda_\mathrm{dr})(\nu-\nu_\mathrm{dr})$ terms and higher-order terms, to motivate a dispersion relation 
\[
 D(\lambda,\nu)=(\lambda-\lambda_\mathrm{dr})^2-(\nu-\nu_\mathrm{dr})^2=((\lambda-\lambda_\mathrm{dr})-(\nu-\nu_\mathrm{dr}))\cdot ((\lambda-\lambda_\mathrm{dr})+(\nu-\nu_\mathrm{dr})),
\]
which, with $\lambda_\mathrm{dr}=0$,  is realized (for instance) in the equation
\begin{align}
 u_t=& +u_x- \nu_\mathrm{dr} u + \beta v,\nonumber\\
 v_t=&-v_x +\nu_\mathrm{dr} v.\label{e:cpw0}
\end{align}
Note that we alternatively could have added a coupling term $\beta u$ in the $v$-equation, but not both terms simultaneously. Integrating \eqref{e:cpw0} on the real line is simple. The $u$ component is advected to the left while decaying exponentially. At the same time $v$ advects to the right while growing exponentially. Compactly supported initial conditions therefore decay to zero in finite time in the absence of the coupling term, $\beta=0$. With the coupling term, however, the $u$-equation integrates an exponentially growing source term stemming from the $v$-equation. To understand the instability, we place an initial condition in $v$ at $x=0$, which is then advected to $x=T/2$ at time $T/2$, with an exponential growth $\rme^{T/2}$. There, it acts as a source term in the $u$-equation, equivalent to an initial condition after some finite time, and transported back in the $u$-equation to $x=0$ with an exponential dampening $\rme^{-T/2}$, yielding in summary the predicted neutral decay $\lambda_\mathrm{dr}=0$. Note that for increasingly long times, this neutral stability requires a sufficiently large domain ahead of the location of observation to enable the reentry of information from $u$ initial conditions. This suggests that arbitrarily distant boundary conditions could eventually impede the instability mechanism.

In our system \eqref{e:cpw}, we set $\nu_\mathrm{dr}=-1$ and introduced a parameter $\mu$ through the transport speed in the $u$-equation. Slower transport in the $u$-equation for $\mu>0$ then increases time for exponential growth and therefore drives an instability in the coupled system. The nonlinear term in the $u$-equation represents generic nonlinear saturation.

\subsection{Resonant spreading speeds and onset of instability for $x\in\R$.}
Detailed criteria for spreading speeds and onset of pointwise instability mediated by resonances were developed in \cite{faye17}. We state the criteria in the relevant scenarios considered here and determine onset of pointwise instability.

\textbf{Onset of pointwise instability --- counter-propagating waves and linear coupling.}
Double roots of the dispersion relation solve 
\[
 D^u(\lambda,\nu)=(1-\mu)\nu+1-\lambda=0,\quad D^v(\lambda,\nu)=-\nu -1-\lambda=0,
 \]
which gives
\[
 \lambda_\mathrm{dr}=\frac{\mu}{2-\mu},\qquad \nu_\mathrm{dr}= -\frac{2}{2-\mu}.
 \]
The double root is pinched when $\mu<1$ and unstable when $\mu>\mu_\mathrm{res}=0$.

\textbf{Onset of pointwise instability --- counter-propagating waves and nonlinear coupling.}
With a source term $v^2$ instead of $v$ in the $u$-equation, the mode in the $u$-equation generated by the coupling is twice the $v$-mode, such that we need to solve 
\[
 D^u(\lambda_1,\nu_1)=(1-\mu)\nu_1+1-\lambda_1=0,\quad D^v(\lambda_2,\nu_2)=-\nu_2 -1-\lambda_2=0, \quad \lambda_1=2\lambda_2,\quad \nu_1=2\nu_2, 
 \]
which gives 
\[
 \lambda_{1,\mathrm{res}}=\frac{2 \mu-1}{2 - \mu},\qquad  
 \nu_{1,\mathrm{res}}=-\frac{3 }{2 - \mu},
\]
with instability for $\mu>\mu_\mathrm{res}=1/2$. The resonance is pinched when $\mu<1$.
 

\textbf{Onset of pointwise instability --- predictions for KPP-KPP.}

We consider \eqref{e:kppkpp} with $d>1$, $\mu>\rho$, and $4d\rho<1$, where the latter condition guarantees pointwise stability of the $v$-equation. The equation $D^u=0,\ D^v=0$ gives in this case two solutions, only one of which is relevant in the sense of \cite{holzerscheel14},
\[
 \lambda_\mathrm{dr}=\frac{
 (d\mu - \rho)  -\sqrt{(d-1)(\mu - \rho)} }{d-1}, 
\]
with instability for 
\[
 \mu>\mu_\mathrm{res}=
\frac{1}{2 d^2}\left(d-1 +2 d \rho + (d-1)\sqrt{(1- 4 d \rho ) }\right).
\]
Below, we will choose $d=5$ which leads to resonant instabilities for $\rho\in(0,\frac{1}{20})$ at $\mu_\mathrm{res}=\frac{1}{50}\left(4+4\sqrt{1-20\rho}+10\rho\right)$. Specifying further, the resonance causes an instability for $\mu\in (0.16,0.25)$ at $\rho=0$. 

In case of quadratic coupling, one needs to solve 
\[
 D^u(\lambda_1,\nu_1)=0, \quad D^v(\lambda_2,\nu_2)=0, \quad \lambda_1=2\lambda_2,\quad \nu_1=2\nu_2.
\]
The resonance condition changes to, assuming $d>2$ and $\mu>2\rho$, 
\[
 \lambda_{1,\mathrm{res}}=\frac{1}{d-2} \left( d \mu +  \sqrt{2(d-2 )(\mu - 2\rho)} - 4  \rho\right), \qquad
 \nu_{1,\mathrm{res}}= \sqrt{\frac{\mu - 2 \rho}{2 (d-2)}}.
\]
At $d=5$, $\rho=0$, the onset of the 2:1-resonant instability occurs for $\mu>\mu_\mathrm{res}=\frac{6}{25}=0.24$, just below the value $\mu=0.25$ where the pinched double root in the $u$-equation becomes unstable.

\textbf{Onset of pointwise instability --- predictions for KPP-CGL.}
In this case, since double roots are complex, the full resonance condition from \cite{faye17} including complex group velocities is needed. 
This resonance condition is 
\[
 D^u(\lambda_1,\nu_1)=0,\   \quad D^A(\lambda_2,\nu_2)=0,  \quad D^{\bar{A}}(\lambda_3,\nu_3)=0, \quad \lambda_1=\lambda_2+\lambda_3,\quad \nu_1=\nu_2+\nu_3,\quad  c_{\mathrm{g},2}=c_{\mathrm{g},3},
\]
where the condition on group velocities $c_{\mathrm{g},j}=-\frac{\rmd \lambda_j}{\rmd\nu_j}$ ensures interaction and optimality \cite{faye17}.
We find for the resonant growth rates, assuming $\mu>2\rho$ and $(1+\alpha^2)d>2$,  
\[
 \lambda_{1,\mathrm{res}}=\frac{ (1+\alpha^2)d\mu - 4 \rho - \sqrt{2 (\mu-2\rho)((1+\alpha^2)d-2)}}{(1+\alpha^2)d-2},\qquad \nu_{1,\mathrm{res}}=-\sqrt{\frac{2(\mu-2\rho)}{(1+\alpha^2)d-2}}.
\]
Instabilities $\lambda_{1,\mathrm{res}}>0$ occur for 
\[
 \mu>\mu_\mathrm{res}=\frac{-2+(1+\alpha^2)d(1+4\rho)+((1+\alpha^2)d-2)\sqrt{1-4(1+\alpha^2)d\rho}}{1+\alpha^2d^2},
\]
where we also assumed $4(1+\alpha^2)d\rho<1$ to guarantee pointwise stability of the CGL component. 

\subsection{Resonant instabilities in large bounded domains and predictions for $\Xi(\mu)$}\label{s:analysis}
The onset of instability discussed above involves interaction of evanescent modes at large distances ahead of the front, as we discussed in \S\ref{s:mod}. In large bounded domains, the boundary conditions limit the distance at which such an interaction can take place and, as a result, suppress the instability partly. In fact, with homogeneous Dirichlet conditions at both $x=-L$ and $x=0$, the trivial solution is stable in the regime where resonant fronts mediate the instability as one can readily infer from the fact that the $v$-component converges to zero and, due to the coupling structure, the $u$-component is then stable as well. We now present an analysis of boundary layer movement in the CPW case, predicting in particular $\Xi(\mu)$ for large $L$, and adapt the results to the more complicated models. 

\begin{figure}
 \includegraphics[width=0.3\textwidth]{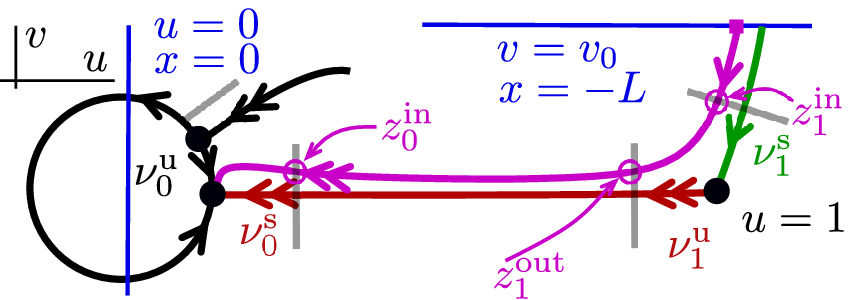}\hfill
 \includegraphics[width=0.3\textwidth]{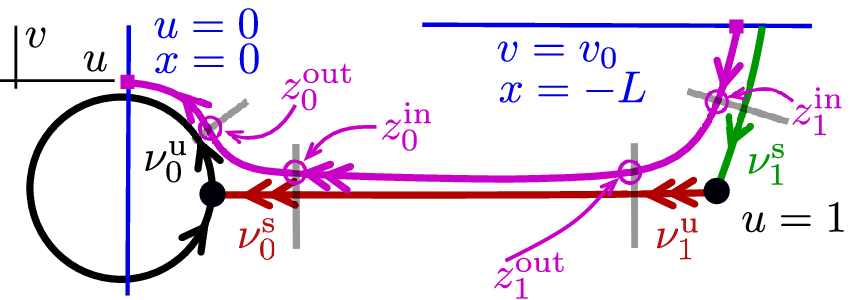}\hfill
 \includegraphics[width=0.3\textwidth]{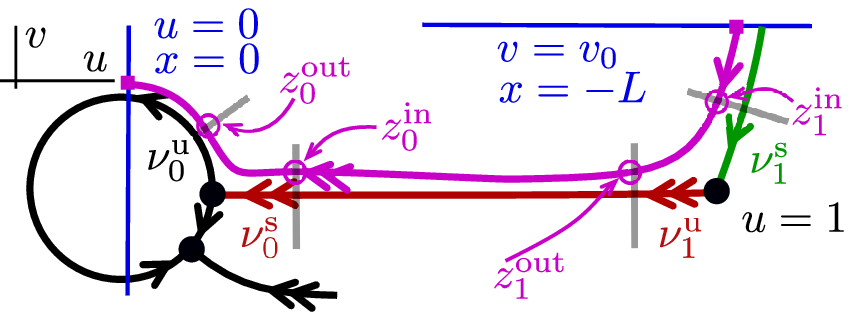}
 \caption{Phase portraits for \eqref{e:cpwphpl} with origin blown up to finite-size circle in polar coordinates at $\mu<0$, $\mu=0$, and $\mu>0$ (left to right). The unbranched resonance manifests as a transcritical bifurcation. Also shown are boundary conditions (blue) and solutions (magenta), blocked for $L$ large at $\mu<0$, with long passage time near equilibria $u=0$ and $u=1$, respectively; see text for analysis of flight times in terms of eigenvalues near equilibria and first-hit maps between sections to the flow near equilibria. Solutions are composed of a heteroclinic boundary layer (green) and a trivial heteroclinic $v\equiv 0$ (red). 
 }\label{f:cpwphpl}
\end{figure}

\begin{figure}
 \centering
 \includegraphics[width=0.3\textwidth]{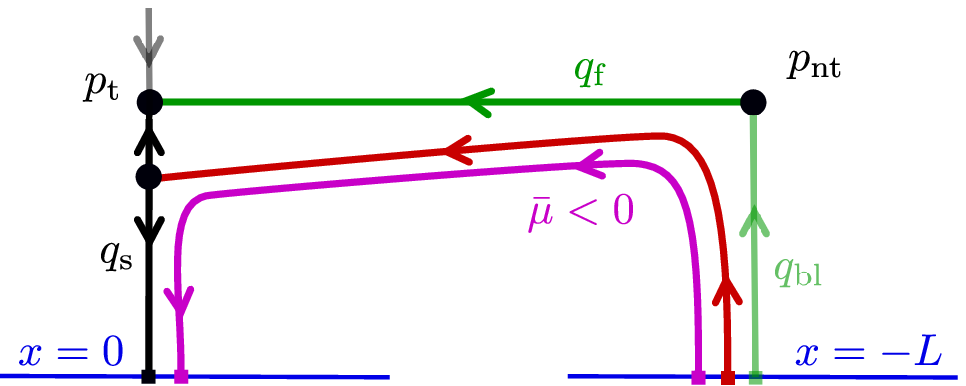}\hspace*{.2in}
 \includegraphics[width=0.3\textwidth]{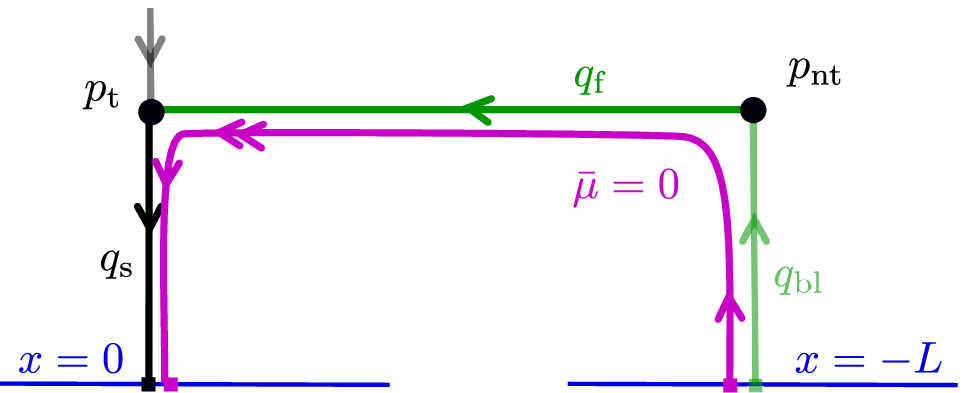}\hspace*{.2in}
 \includegraphics[width=0.3\textwidth]{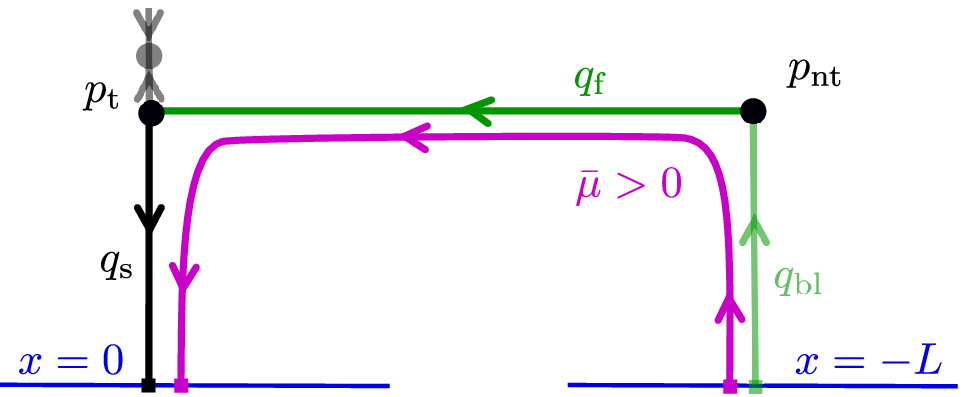} 
 \caption{Illustration of the heteroclinic chains responsible for the bifurcation diagram below (left) at (center) and above (right) criticality. Below criticality, the stable manifold of the bifurcated equilibrium provides upper bounds on the passage time near $p_\mathrm{nt}$ and thereby $\Xi(\mu)=0$ for $L\to\infty$. At criticality, the expansion near $p_\mathrm{t}$ is weak, algebraic, leading to long passage times near $p_\mathrm{t}$ and therefore small upper bounds on the passage times near $p_\mathrm{nt}$. For $\bar{\mu}>0$, passage times are balanced by eigenvalue ratios as explained in the text. }\label{f:cpw_schem}
 \end{figure}

\textbf{Predicting $\Xi(\mu)$ for counter-propagating waves with linear coupling.}
We look for steady-states of \eqref{e:cpw} solving 
\begin{equation}\label{e:cpwphpl}
 u_x=\frac{1}{1-\mu}\left(-u+u^3-v\right),\qquad v_x=-v.
\end{equation}
A phase plane analysis shows 3 equilibria, $u\in \{-1,0,1\}$ and $v=0$, with eigenvalues $\{-\frac{1}{1-\mu},-1\}$ at $(0,0)$ and $\{\frac{2}{1-\mu},-1\}$ at $(\pm1,0)$.  In order to resolve the double eigenvalue at the origin for $\mu=0$, we use polar coordinates as in the case of the branched resonance. The phase portrait is shown in Fig.~\ref{f:cpwphpl}. Total flight time near boundary conditions is $L=T_0+T_1+\rmO(1)$, where $T_0$ and $T_1$ are the flight times between sections to the flow near the equilibria $u=0$ and $u=1$, respectively. 
The heteroclinic shown in red is found from the heteroclinic solution to $u_x=-u+u^3$, $u(x)\to 1$, for $x\to -\infty$, and $u(x)\to 0$ for $x\to\infty$. The green boundary layer is found from integrating the stable manifold of $u=1$ backward in time using that $v$-dynamics are decoupled. A schematic representation of the heteroclinic chains is found in Fig.~\ref{f:cpw_schem}.

Flight times can be computed from a Shilnikov passage time analysis: near a hyperbolic equilibrium at the origin in the plane $(U_\mathrm{s},U_\mathrm{u})$ with eigenvalues $\kappa^\mathrm{s}$ and $\kappa^\mathrm{u}$, one can find a passage map for instance through linearizing,
\[
 U_\mathrm{s}'=-\kappa^\mathrm{s} U_\mathrm{s},\qquad
 U_\mathrm{u}'=\kappa^\mathrm{u} U_\mathrm{u},
\]
and find that the first-hit map from the in-section 
$\{U_\mathrm{s}=\delta,\ U_\mathrm{u}=U_\mathrm{in}>0\}$ to the out-section 
$\{U_\mathrm{u}=\delta,\ U_\mathrm{s}=U_\mathrm{out}>0\}$ is explicitly given through 
\[
U_\mathrm{out}=\delta^{
(\kappa^\mathrm{u}-\kappa^\mathrm{s})/\kappa^\mathrm{u}
}
U_\mathrm{in}^{\kappa^\mathrm{s} /\kappa^\mathrm{u} }
\text{ with passage time }
T=\log\delta -\frac{1}{\kappa^\mathrm{u}}\log U_\mathrm{in}.
\]
This analysis near the equilibrium $u=1$ gives
\begin{equation}\label{e:sh1}
 z_0^\mathrm{in}\sim z_1^\mathrm{out} \sim \left(z_1^\mathrm{in}\right)^{\nu_1^\mathrm{s}/\nu_1^\mathrm{u}}, \qquad T_1\sim -\frac{1}{\nu_1^\mathrm{u}}\log z_1^\mathrm{in},
\end{equation}
where we neglected constants and higher-order terms induced by linearizing coordinate changes. The eigenvalue $\nu_0^\mathrm{u}=\mu$ gives dynamics in projective space and is given by the difference between the two stable eigenvalues at $u=0$. Assuming $\nu_0^\mathrm{u}=\rmO(1)$, we can use the same simple linear analysis to compute 
\begin{equation}\label{e:sh0}
 z_1^\mathrm{out}\sim \left(z_1^\mathrm{in}\right)^{|\nu_0^\mathrm{s}|/\nu_0^\mathrm{u}}, \qquad T_0\sim -\frac{1}{\nu_0^\mathrm{u}}\log z_0^\mathrm{in}\sim -\frac{|\nu_1^\mathrm{s}|}{\nu_0^\mathrm{u}\nu_1^\mathrm{u}}\log z_1^\mathrm{in}\sim  \frac{|\nu_1^\mathrm{s}|}{\nu_0^\mathrm{u}}T_1
\end{equation}
Together with $L\sim T_0+T_1$, this gives
\begin{equation}\label{e:resfarfromonset}
\Xi(\mu)\sim \frac{\nu_0^\mathrm{u}}{\nu_0^\mathrm{u}+|\nu_1^\mathrm{s}|}.
\end{equation}
Substituting values $\nu_0^\mathrm{u}=-1-\frac{-1}{1-\mu}=\frac{\mu}{1-\mu}$, $\nu_1^\mathrm{s}=-1$ from \eqref{e:cpwphpl} gives 
\begin{equation}
 \Xi(\mu)\sim \mu.\label{e:xicpw}
\end{equation}

For $\mu\sim 0$, linearization at the origin does not give accurate predictions due to quadratic terms in the transcritical bifurcation. 
Including the quadratic term with a coefficient $\beta$ gives a time of flight 
\begin{equation}\label{e:t0cmd}
 T_0=\frac{1}{\nu_0^\mathrm{u}} \log \left( 
 \frac{\nu_0^\mathrm{u}+\beta z_0^\mathrm{in}}{( \nu_0^\mathrm{u}+\beta)z_0^\mathrm{in}} 
 \right)=
 \frac{1}{\nu_0^\mathrm{u}} \log \left(1+ \frac{\nu_0^\mathrm{u}}{\beta z_0^\mathrm{in}} \right)+\rmO(1),\qquad T_1=-\frac{1}{\nu_1^\mathrm{u}}z_1^\mathrm{in}.
\end{equation}
Eliminating $z_0^\mathrm{in}$ using \eqref{e:sh0} and $T_0=L-T_1$ gives 
\[
 L-T_1=\frac{1}{\nu_0^\mathrm{u}} \log\left(1+c\nu_0^\mathrm{u}\rme^{|\nu_1^\mathrm{s}|T_1}\right), 
\]
for some constant $c>0$. This yields an implicit equation for $\Xi$,
\begin{equation}\label{e:Lcmf}
\Xi-1-\frac{1}{\nu_0^\mathrm{u}L} \log\left(1+c\nu_0^\mathrm{u}\rme^{|\nu_1^\mathrm{s}|L\Xi}\right)=0. 
\end{equation}
One can derive more explicit expressions in asymptotic regimes 
\begin{equation} \label{e:smallgap}
 T_1=\left\{
 \begin{array}{ll}
  \frac{\nu_0^\mathrm{u}}{\nu_0^\mathrm{u}+|\nu_1^\mathrm{s}|}L,& \nu_0^\mathrm{u}\rme^{|\nu_1^\mathrm{s}|L\Xi}\gg 1,\\
  \frac{1}{|\nu_1^\mathrm{s}|} \log L + \frac{\nu_0^\mathrm{u}}{2(1+|\nu_1^\mathrm{s}|)}L,& \nu_0^\mathrm{u}\rme^{|\nu_1^\mathrm{s}|L\Xi}\ll 1, 
 \end{array}
 \right.
\end{equation}
Of course, the regime $\nu_0^\mathrm{u}\rho\gg 1$ in  \eqref{e:smallgap} simply recovers \eqref{e:resfarfromonset}. 

We conclude by summarizing the key conceptual elements of the analysis, also illustrated in Fig. \ref{f:cpw_schem}. In fact, the list is almost identical to the list in the case of a branch resonance in \S\ref{s:ap} with two changes:
\begin{itemize}
 \item[(ii)'] a transcritical bifurcation of equilibria on the critical circle reflecting the unbranched resonance in a crossing of eigenspaces. 
 \item[(iv)'] a heteroclinic orbit $q_\mathrm{f}$, the invasion front on the unbounded domain, between nontrivial equilibrium and trivial equilibrium, \emph{contained} in the strong stable manifold of the saddle-node equilibrium at criticality.
\end{itemize}
Of course, since the bifurcation here is nonlocal in parameter space, extending over an interval of $\mu$-values until reaching a branched resonance, one needs to require existence of heteroclinics and equilibria over the entire range of parameter values where predictions are to be derived.  Assumption (iv)' is typical due to the presence of invariant subspaces such as $A=A_x=0$ or $v=v_x=0$ in CGL-KPP and KPP-KPP, respectively. In systems without gauge invariance, one finds such invariant subspaces in spatial-dynamics formulations for time-periodic functions as the subspace of trivial, constant time-dependence; see for instance \cite{ssessfront} for an example illustrating bifurcation from heteroclinic orbits within such subspaces. 

\textbf{Predicting $\Xi(\mu)$ for counter-propagating waves with quadratic coupling.}
The steady-state equation with quadratic coupling reads 
\begin{equation}\label{e:cpwphpl2}
 u_x=\frac{1}{1-\mu}\left(-u+u^3-v^2\right),\qquad v_x=-v.
\end{equation}
One can now directly set $w=v^2$, which gives $w_x=-2w$ and continue the previous analysis with $\nu_1^\mathrm{s}=-2$ and find 
\begin{equation}\label{e:xicpw2}
 \Xi(\mu)\sim 2\mu-1.
\end{equation}
In a more general approach, one would consider a desingularization near the origin using the scaling from the 2:1-resonance, setting $(u,v^2)=R(\cos(\varphi),\sin(\varphi))$ in $\{v>0\}$, with projective coordinates $v^2/u$, in which the transcritical bifurcation occurs at $\mu=1/2$.

\textbf{Predicting $\Xi(\mu)$ for the coupled KPP system.} 
The analysis in this case completely parallels the analysis in the previous case, although the 4-dimensional phase portraits are less accessible. One still finds two relevant equilibria, $u=\sqrt{\mu}$ and $u=0$, $v=0$, and heteroclinic orbits connecting the boundary conditions at $x=-L$ to $u=\sqrt{\mu}$, then $u=\sqrt{\mu}$ to $u=0$ within $v=0$, and a heteroclinic in the singular sphere of the blown up origin. One finds 
\[
 \nu_1^\mathrm{s}=\kappa\frac{-1-\sqrt{1-4 d \rho}}{2d},\qquad 
 \nu_0^\mathrm{u} = \frac{1-\sqrt{1-4 \mu}}{2}+\nu_1^\mathrm{s},
\]
where $\kappa$ denotes the power of the coupling terms and determines the order of the resonances. With these adaptations, one then recovers all formulas from the previous case. 

\textbf{Predicting $\Xi(\mu)$ for the coupled CGL-KPP system.}
With the particular skew-product structure, this situation is very similar to the coupled KPP system. For now, assume that $|A|$ is stationary, that is, $A$ is stationary up to the gauge symmetry, and  let $\nu^1_\mathrm{s}$ be the exponential rate of decay of $|A|$. We then recover all formulas from the case of counter-propagating waves, with $\nu_0^\mathrm{u}=\frac{1-\sqrt{1-4\mu}}{2}+\nu_1^\mathrm{s}$. It turns out that the value of $\nu_1^\mathrm{s}$ depends on the boundary condition through the frequency $\beta$ in a nontrivial fashion. Varying the boundary conditions, we found stationary profiles for $|A|$ with zero and nonzero frequencies $\beta$, but also, in direct simulations, time-periodic solutions. We saw exponential decay rates varying significantly changing only boundary conditions. We measured the exponential decay rate numerically and also compared with rates obtained from the dispersion relation when the boundary layer is stationary up to the gauge symmetry and used the result to obtain $\nu_1^\mathrm{s}$ and predictions for the bifurcation diagram.

\begin{figure}[h]
 \includegraphics[width=0.3\textwidth]{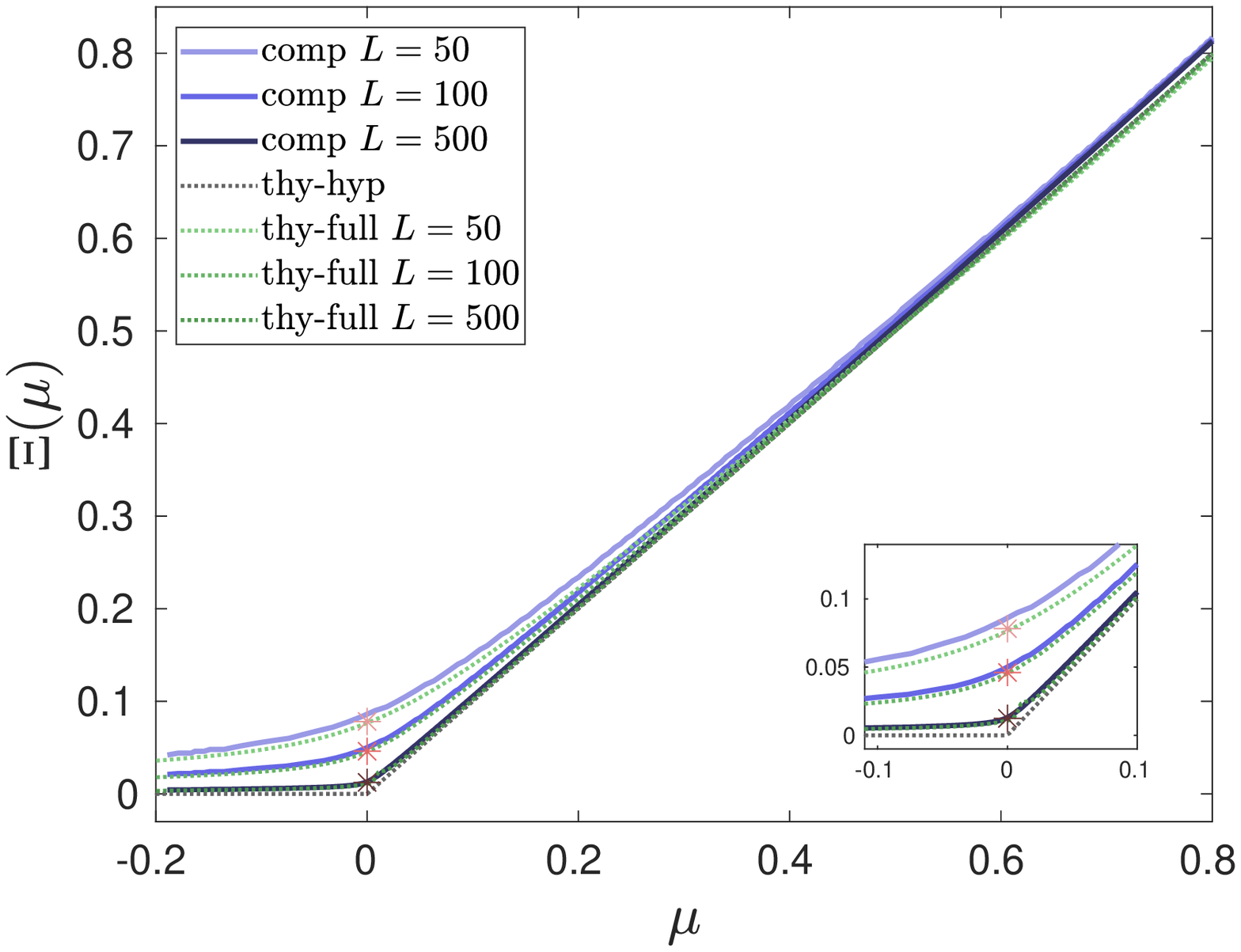}\hfill\includegraphics[width=0.3\textwidth]{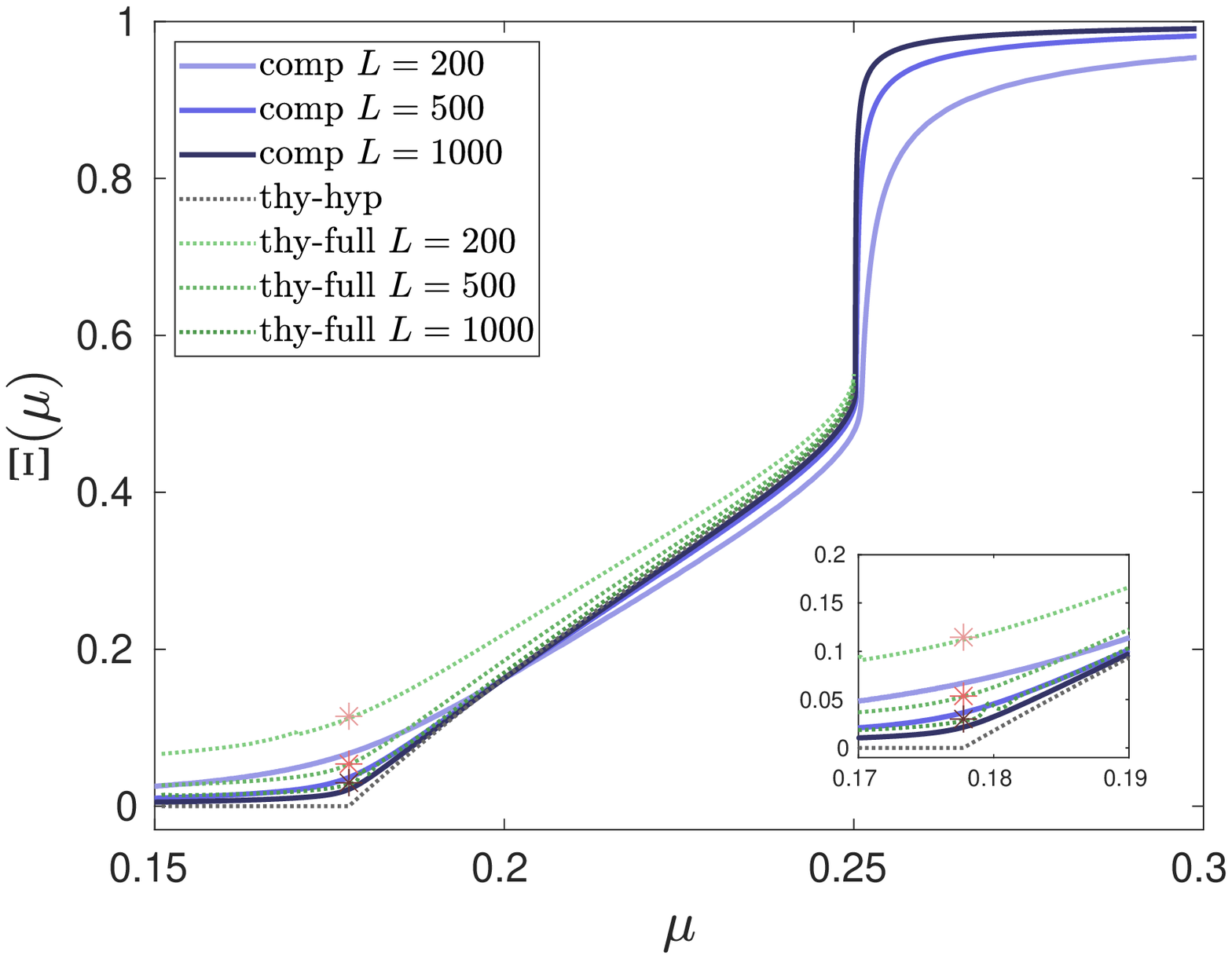}\hfill \includegraphics[width=0.3\textwidth]{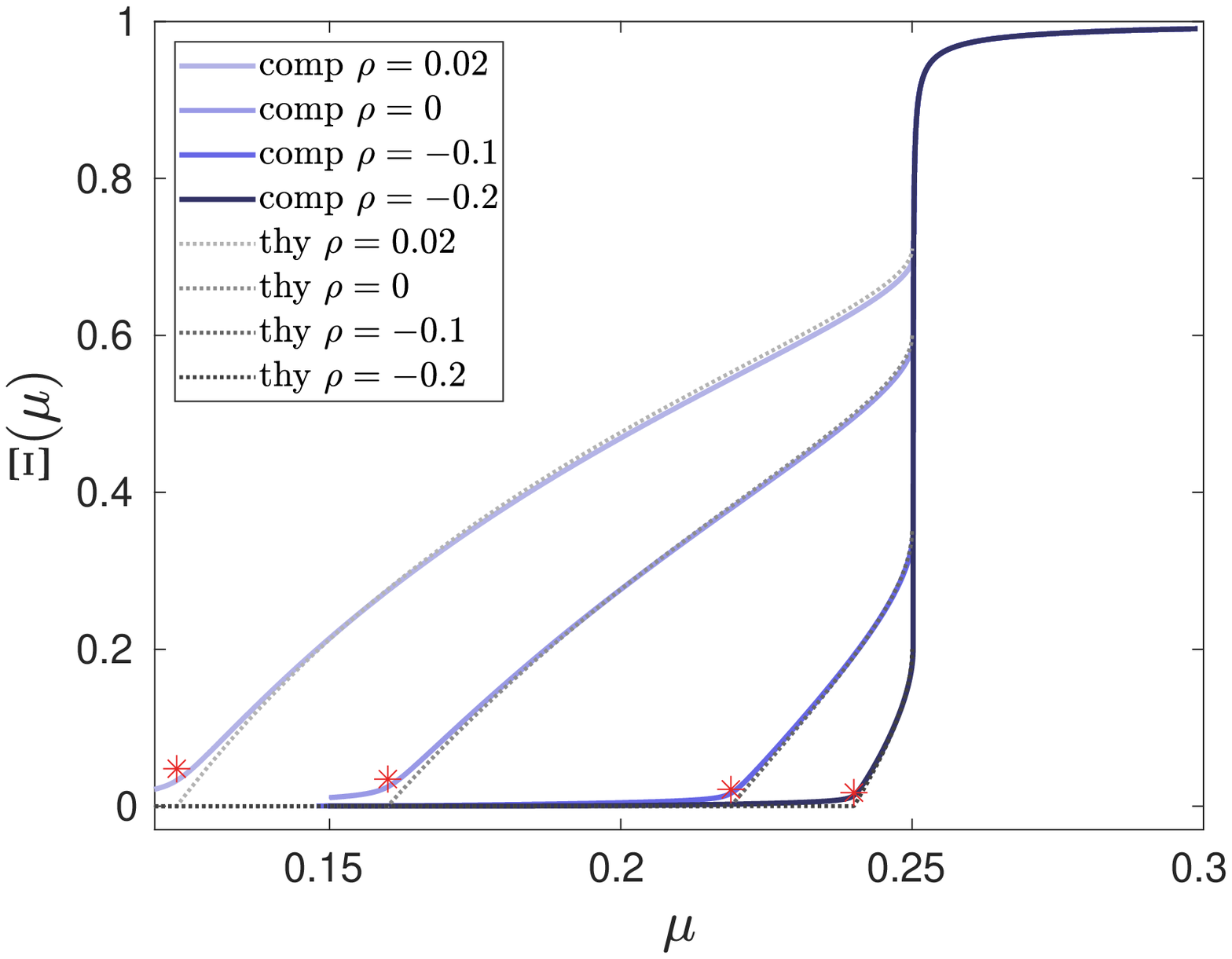}
  \caption{Comparison of $\Xi(\mu)$ from numerical continuation to theory. Counter-propagating wave system (\eqref{e:cpw}, left) for $L=50,100,500$ and coupled KPP system (\eqref{e:kppkpp}, center), $\rho=20.5^{-1}$, $\kappa=2$, $u_0=0.5$, $v_0=0.1$) for $L=200,500,1000$;  comparison with theory \eqref{e:xicpw} (gray, dotted) and \eqref{e:Lcmf} (green, dotted, $c=1$); approximation at $\mu=\mu_\mathrm{res}$ from \eqref{e:Lcmf} marked with ``$*$''. Bifurcation curves with $L=1000$, $d=10$, varying $\rho$, same parameters as before (right).  }\label{f:res}
\end{figure}

\subsection{Numerical continuation, comparison with theory}\label{s:comp}
We computed bifurcation diagrams numerically via secant continuation and in direct simulations. Comparisons with theoretical predictions are shown in Fig.~\ref{f:res}.  Exponential decay is stronger in \eqref{e:cpw} compared to the numerical experiments for KPP and CGL, such that domain sizes are comparatively smaller in the plots. Bifurcation diagrams depend little on the boundary conditions but change depending on parameters such as $\rho$. Note that one eventually always sees the branched resonance set in and dominate the eventual climb to $\Xi=1$ in Fig.~\ref{f:res} for the KPP-KPP and CGL-KPP system (but of course not for the CPW system, which does not have a branched resonance but becomes ill-posed when $\mu=\Xi(\mu)=1$), consistent with the fact that $\Xi<1$ in \eqref{e:resfarfromonset}.

\begin{figure}
 \includegraphics[width=0.3\textwidth]{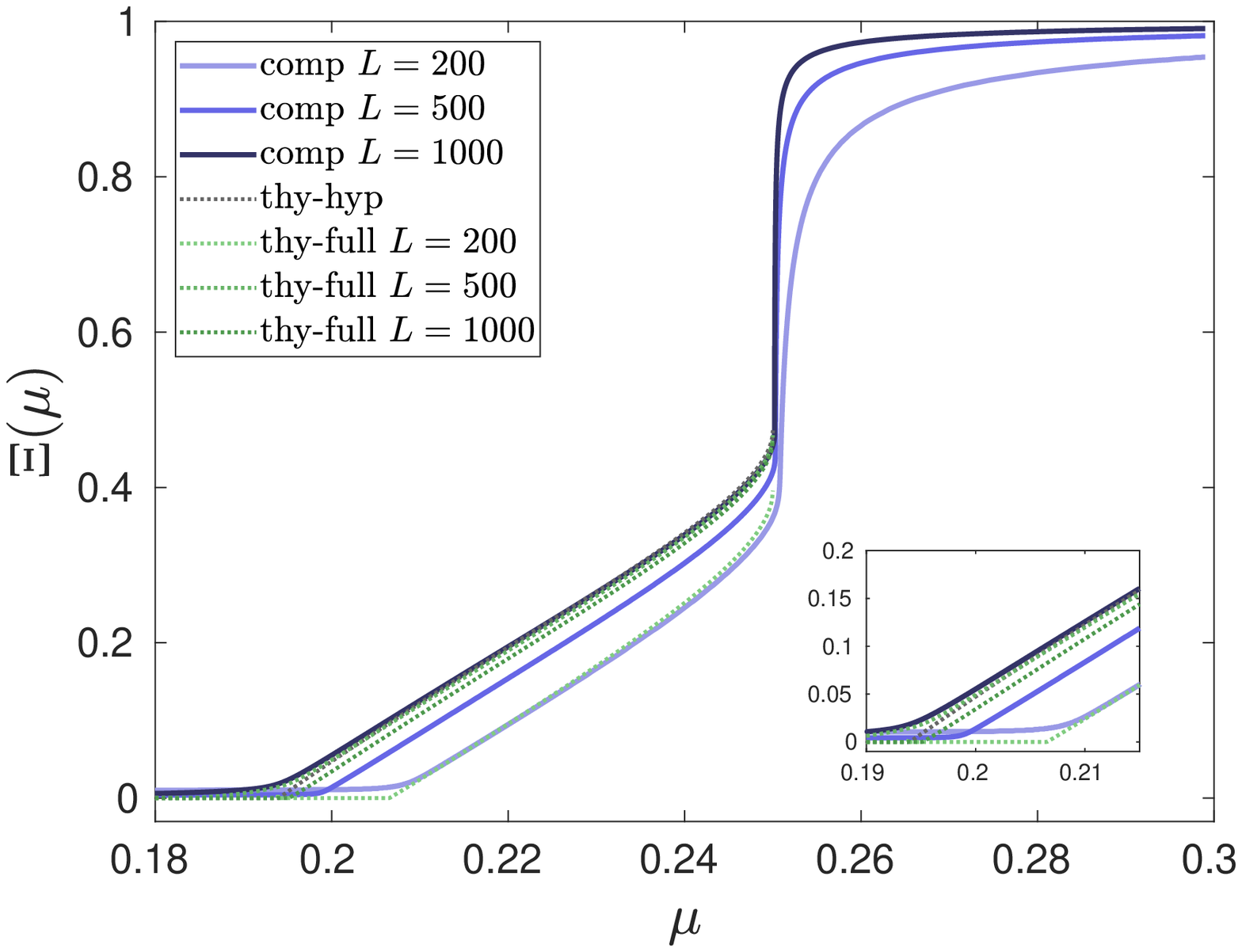}\hfill
 \includegraphics[width=0.3\textwidth]{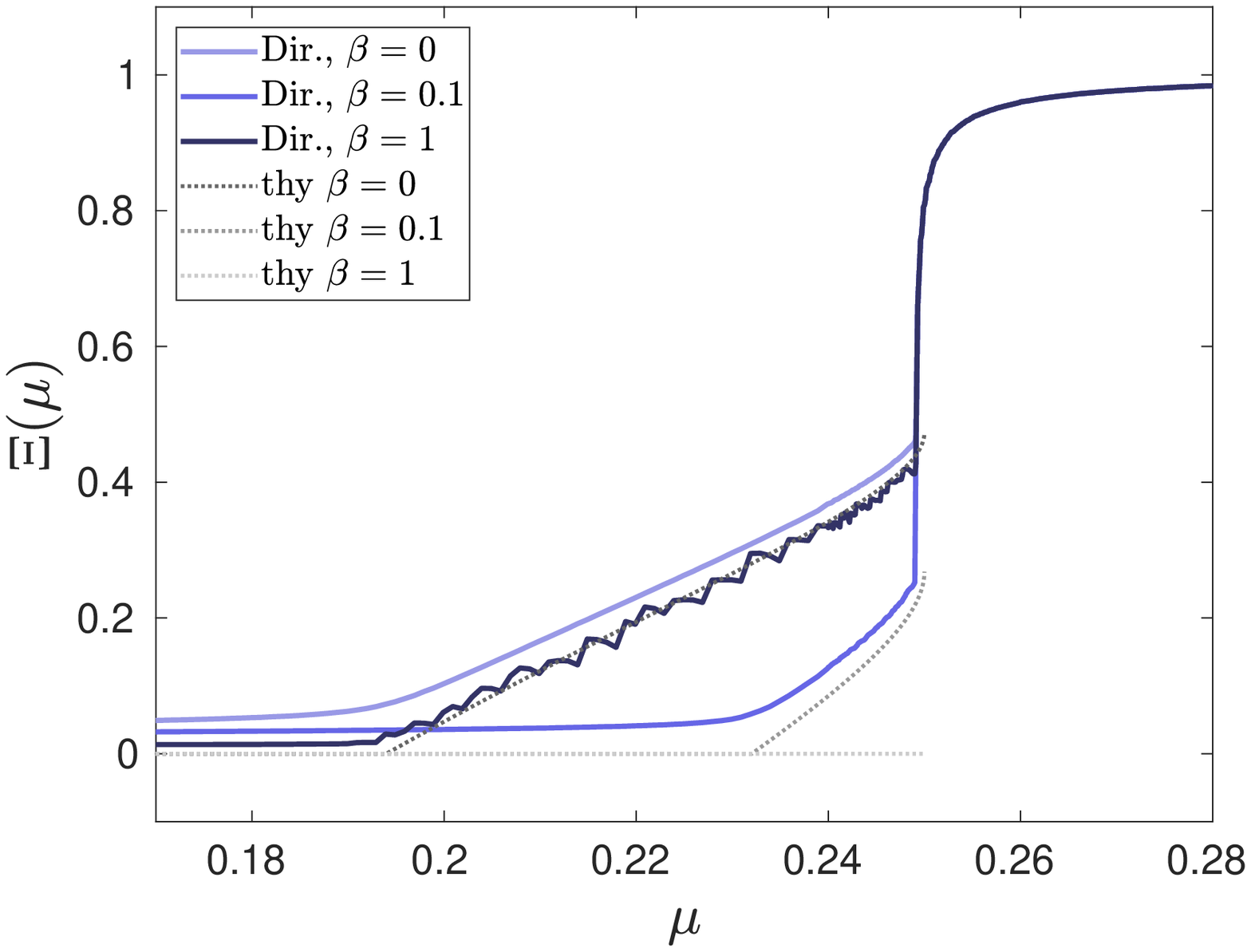}\hfill
 \includegraphics[width=0.3\textwidth]{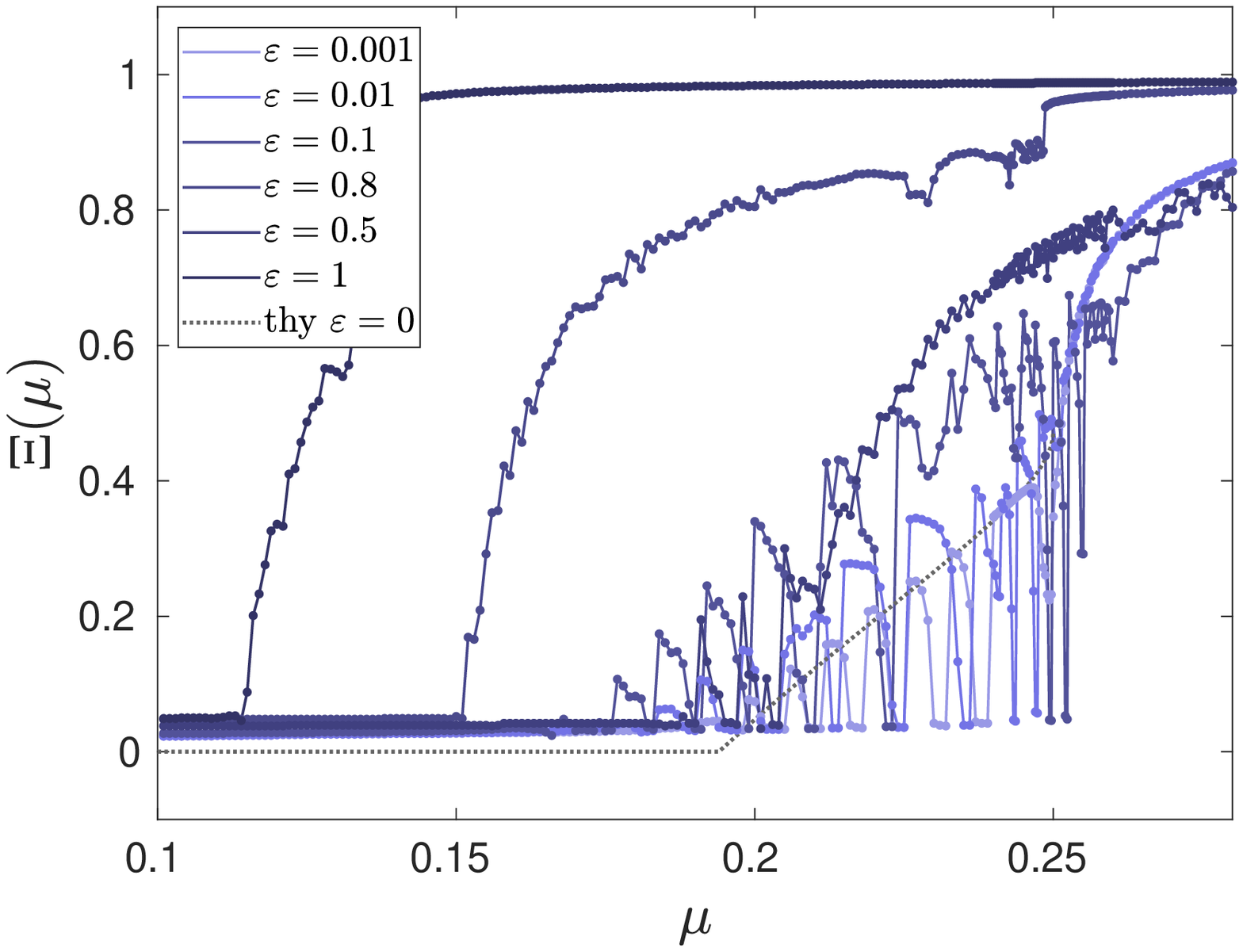}
  \caption{Comparison of $\Xi(\mu)$ from numerical continuation to theory in the CGL-KPP system \eqref{e:kppcgl}, parameters $\alpha=0.3,\,\beta=0,\, \gamma=0.3,\, d=5,\,\rho=0.04,\, R_0=0.04,\,k_0= 0.5$ (left). Data from direct simulations, $L=500$, same parameter values, varying $\beta$ (center),  and introducing back coupling \eqref{e:kppcglback} for $\beta=1$, varying $\eps$, $L=100$, with predictions for $\eps=0$. 
 }\label{f:rescglkpp}
\end{figure}

For the coupled CGL-KPP system \eqref{e:kppcgl} and gauge-invariant boundary conditions, we computed the solution to CGL using Newton's method with an initial guess from direct simulations, substituted it into the secant-continuation algorithm for the KPP subsystem,  and measured its exponential decay rate for use in the theoretical predictions. The CGL solution turns out to be very small for our choices of parameters $R_0=0.04, k_0=0.5$, $\alpha=0.3,\gamma=0.3$, with amplitude $10^{-3}$. As a consequence, fits are less good for finite $L$. Roughly adjusting the constant $c$ in \eqref{e:Lcmf} to $c=10^4$ for $L=200,500$ and $c=10$ for $L=1000$ does yield quite good fits, nevertheless; see left panel in Fig. \ref{f:rescglkpp}. We used the fact that the boundary layer was stationary in time to obtain the exponential decay rate from the dispersion relation with $\lambda=0$, with excellent agreement to measured exponential decay. 

We also compared predictions with direct simulations and Dirichlet boundary conditions $A_0=0.6$ and $\beta=0,0.1,1$; see Fig. \ref{f:rescglkpp}, center panel. For each $\beta$, one can calculate an exponential decay rate of a stationary boundary layer from the stationary ODE, which yields $\nu_1^\mathrm{s}$ and thereby predictions for the position $\Xi(\mu)$. The exponential decay rate matches numerical observations and predictions give good fits for moderate values of $\beta$. For large $\beta$, the boundary layer oscillates, and generates an effective decay  similar to the one for a stationary boundary layer with frequency close to zero. Predictions for a boundary layer with frequency $\beta$ would predict absence of a resonant instability but oscillations in the boundary layer mediate a weaker decay that drives a resonant instability, albeit not quite as strong as in the case $\beta=0$. In fact, $\beta=-\alpha\rho=-0.12$ would yield weakest decay, only slightly less than in the case $\beta=0$. Position measurements for $\beta=1$ are instantaneous in time and therefore vary slightly depending on the phase of the oscillation.

Lastly, we demonstrate the robustness of the phenomenon described here by introducing back coupling, that is, considering 
\begin{align}
  u_t=&\  u_{xx} +u_x + \mu u - u^3 + |A|^2 +\eps \,\Re\,A, \quad 0<x<L,\nonumber\\
  A_t=&\ d(1+\rmi\alpha)A_{xx}+A_x + (\rho+\rmi\beta) A - (1+\rmi\gamma)A|A|^2 + \eps u, \quad 0<x<L,\label{e:kppcglback}
\end{align}
with Dirichlet boundary conditions $A_0=0.6$ and $u=0.1$, $\eps=0.001,0.01,0.1,0.5,0.8,1.0$, increasing the strength of interaction, with $\beta=1,\ \rho=0.04$, and forcing oscillations in the boundary layer; see Figure \ref{f:rescglkpp}, right panel. We see that  ``bifurcation'' curves are roughly continuous in $\eps$, with $\Xi$ increasing in $\eps$. Dynamics are chaotic, and we measured instantaneous values of the position, but waiting for decrease in $\Xi$ to stabilize, leading to somewhat erratic plots with resulting upward jumps. The possibility of intermittent collapse of the interface, that is, $\Xi(\mu)$ dropping to near-zero values for some ranges of parameter values, does appear to be robust however with respect to changes in the numerical algorithms and homotopy strategy. Since exponentially weak interactions are responsible for all phenomena here, it is however difficult to quantify the effect of round-off errors without strong theoretical predictions as available in the case of stationary solutions.  For stronger coupling, $\eps>0.5$, a branched resonance determines the onset of instability.

\section{Discussion}\label{s:d}
We investigated instabilities in large domains in the presence of unidirectional transport. Instabilities are triggered when the speed of propagation of large, nonlinear fronts reverses sign, and a front propagates into the domain from the downstream boundary. We focused on cases where this front propagation is determined by the dispersion relation of the trivial state, that is, on supercritical instabilities leading to pulled fronts. The speed of such pulled fronts is most commonly identified with a branch point of the dispersion relation, and bifurcation diagrams in large domains universally exhibit a very steep onset with square root-like asymptotics for the position of the interface. Our main contribution identifies an alternate, robust scenario, where the front speed in the unbounded domain is determined by a nontrivial, unbranched resonance. Bifurcations in large domains are in this case gradual in that the position of the front interface as a fraction of the domain size increases gradually. The cause of the arrest of the front far away from boundaries is a cut-off in tails caused by the upstream boundary, which suppresses the key resonance mechanism. In both cases, our expansions can be thought of as identifying corrections to an unbounded-domain approximation and thereby quantifying in which sense $L$ is large, approximating an unbounded domain. 

More directly, an immediate consequence of our results are estimates for the effect of boundaries on dynamics in large domains in the presence of transport. Experimental or numerical results for the onset of instability in finite  domains need to be corrected when making predictions for infinite or different size domains. Those corrections are negligible in the case of pushed (or bistable) fronts, where a standard heteroclinic bifurcation analysis predicts a steep transition in a parameter window that is exponentially small. For pulled fronts and branched resonances, corrections are small, $\rmO(L^{-2})$ but not negligible. For unbranched resonances, corrections are in fact $\rmO(1)$! The simplest manifestation of this notable effect arises when attempting to compute invasion speeds using a Newton-type freezing method, where one chooses artificial boundary conditions in a small, finite-size domain, fixes the position of an interface through a phase condition, and adds the speed as a Lagrange multiplier; see for instance \cite{beynfreeze,stegemerten}. 
Our results show that the error in the speed $\delta c$ resulting from the truncation to a finite domain is 
\begin{itemize}
    \item $\delta c =\rmO(\rme^{-\eta L})$ for pushed fronts;
    \item  $\delta c =\rmO(L^{-2})$ for pulled fronts in branched resonances;
    \item  $\delta c =\rmO(1)$ for pulled fronts in unbranched resonances.
\end{itemize}
Since fronts decay exponentially in the leading edge, $L$ usually needs to be moderate such that the leading edge can be resolved numerically, which, depending on algorithms and machine precision limits $L$ usually to $100-1000$ for decay rates $\rme^{-L}$, which leads to non-negligible errors in speed predictions even in the branched case. These errors should be compared to errors $\rmO(t^{-1})$ in the speed in direct temporal simulations due to the logarithmic shift in positions of pulled fronts in the branched case \cite{avery2020universal,Bramson2,Comparison1}. Temporal corrections are not known in the unbranched case. 

Our bifurcation analysis is  based on a conceptual representation of boundary layers and front interfaces as chains of heteroclinic orbits. A key step in the analysis is a geometric desingularization at the origin which reduces eigenvalue resonances to saddle-node and transcritical bifurcations of eigenspaces in the branched and unbranched case, respectively, and introduces an additional singular heteroclinic orbit connecting eigenspaces and boundary conditions into the geometric picture of heteroclinic chains. Since such tail interaction and heteroclinic chain bifurcations are notoriously difficult to analyze in general, relying on multiple non-degeneracy and non-resonance assumptions to quantify exponential expansions, it seems cumbersome to state general results rigorously. We do believe however that the situation analyzed here is in many ways generic and hope that a  rigorous dynamical systems analysis of the heteroclinic bifurcation in a more general setup could clarify such genericity assumptions. 

In direct simulations, small initial conditions lead to fronts propagating into the domain until changing speed and settling the position of the interface at the predicted location. We intend to study transient front dynamics in large and semi-bounded domains in future work. 

In a different direction, weak tail interaction is sensitive to noise, additive or mutiplicative. It was demonstrated in \cite{holzerremnant} that multiplicative noise changes the resonance condition and absolute spectra \cite{ssabs} then determine changed, faster spreading speeds. It would be interesting to analyze such effects in the case of large bounded domains, as well. 

\def\cprime{$'$}

%
%

\end{document}